\documentclass[reprint,superscriptaddress,pra,nofootinbib]{revtex4-2}

\usepackage{graphicx}    % Graphics handling of JPEG, PNG, PDF
\usepackage{microtype}   % Finetuned typesetting control
\usepackage{amsmath}     % AMS Math package; equation environments, math addons
  \usepackage{amsthm}    % AMS theorem-like environments
  \usepackage{amssymb}   % AMS math symbols and styles like \mathbb/cal/frak/... (loads amsfonts)
\usepackage{siunitx}     % Unified number/unit handling
\usepackage{booktabs}    % Better table control
\usepackage[colorlinks,citecolor=blue]{hyperref}
\usepackage[capitalize]{cleveref}    % Smart referencing
  \crefname{algocf}{Alg.}{Algs.}
  \Crefname{algocf}{Algorithm}{Algorithms}
  \crefname{section}{Sec.}{Secs.}
  \crefname{appendix}{App.}{Apps.}

\usepackage{adjustbox}    % Circuit size control
\usepackage{subcaption}
\usepackage{stackengine}
\usepackage{xcolor}
\usepackage{mathtools} % Fine-tune math typesetting
\usepackage{bm}
\usepackage[shortlabels]{enumitem}  % Customize list environments
\usepackage[ruled,vlined]{algorithm2e}
\usepackage{tikz}
  \usetikzlibrary{math}
  \usetikzlibrary{decorations.pathreplacing}
  \usetikzlibrary{quantikz}

\providecommand{\ord}{O}
\providecommand{\ie}{\emph{i.e.}}
\providecommand{\eg}{\emph{e.g.}}

\let\originalleft\left
\let\originalright\right
\renewcommand{\left}{\mathopen{}\mathclose\bgroup\originalleft}
\renewcommand{\right}{\aftergroup\egroup\originalright} % Fixes spacing for resized delimiters

\DeclarePairedDelimiter{\ceil}{\lceil}{\rceil}

\DeclarePairedDelimiter\norm{\lVert}{\rVert}

\DeclareMathOperator{\erf}{erf}
\providecommand{\nnl}{\nonumber \\}

\newcommand{\Z}{\mathbb{Z}}

\begin{document}

\title{Practical considerations for the preparation of\texorpdfstring{\\}{ }multivariate Gaussian states on quantum computers}

\author{Plato Deliyannis\texorpdfstring{$^{\P}$}{}}
\email{pdeliyannis@lbl.gov}
\affiliation{Physics Division, Lawrence Berkeley National Laboratory, Berkeley, CA 94720, USA}

\author{Marat Freytsis\texorpdfstring{$^{\P}$}{}}
\email{marat.freytsis@rutgers.edu}
\affiliation{NHETC, Department of Physics and Astronomy, Rutgers University, Piscataway, NJ 08854, USA}
\affiliation{Physics Division, Lawrence Berkeley National Laboratory, Berkeley, CA 94720, USA}

\author{Benjamin Nachman}
\email{bpnachman@lbl.gov}
\affiliation{Physics Division, Lawrence Berkeley National Laboratory, Berkeley, CA 94720, USA}

\author{Christian W. Bauer}
\email{cwbauer@lbl.gov}
\affiliation{Physics Division, Lawrence Berkeley National Laboratory, Berkeley, CA 94720, USA}

\date{\today}

\begin{abstract}
  We provide explicit circuits implementing the Kitaev--Webb algorithm~\cite{kitaev2009wavefunction} for the preparation of multi-dimensional Gaussian states on quantum computers.
  While asymptotically efficient due to its polynomial scaling, we find that the circuits implementing the preparation of one-dimensional Gaussian states and those subsequently entangling them to reproduce the required covariance matrix differ substantially in terms of both the gates and ancillae required.
  The operations required for the preparation of one-dimensional Gaussians are sufficiently involved that generic exponentially-scaling state-preparation algorithms are likely to be preferred in the near term for many states of interest.
  Conversely, polynomial-resource algorithms for implementing multi-dimensional rotations quickly become more efficient for all but the very smallest states, and their deployment will be a key part of any direct multidimensional state preparation method in the future.
\end{abstract}

\maketitle
\def\thefootnote{\P}
\footnotetext{These authors contributed equally.}
\def\thefootnote{\arabic{footnote}}

\section{Introduction}
\label{sec:intro}
The simulation of real-world physical systems whose quantum effects are difficult to capture using classical resources has been a natural application for quantum computers since their beginning~\cite{Feynman:1981tf} and is a key goal for the current noisy intermediate-scale quantum computing (NISQ) era~\cite{Preskill_2018}.
The initial step of any such simulation requires the preparation of a physical state in a representation suitable for the available computing paradigm.
Similar considerations apply for final state preparation often required for readout. 
Especially for problems involving real-time dynamics, where sign problems typically plague both classical and quantum sampling methods, good approximations to fill states are an important part of any simulation pipeline.
A natural approach to this task is to explicitly encode a known wavefunction on some number of qubits, potentially supplemented with an evolution to a more difficult to access state.
The need for efficient state preparation algorithms has been recognized for a long time~\cite{Zalka:1996st,Kaye:01,Grover:576789,soklakov2004efficient}, and this topic continues to be an active area of research including adiabatic~\cite{Lamm:2018siq,Harmalkar:2020mpd} and variational~\cite{Kaplan:2017ccd,Arrazola_2019,Liu:2021otn} methods.

A general scheme for the preparation of one-dimensional real wavefunctions was proposed in~\cite{Zalka:1996st} (and rediscovered in~\cite{Kaye:01,Grover:576789}).
It involves recursively building up states in a qubit register starting from the most-significant qubit by rotations whose angles are controlled by the values of qubits already set.
Such a scheme is efficient provided the required angles, which correspond to integrating the wavefunction probability density over arbitrary intervals, can themselves be computed efficiently.
This is true for, \eg, any log-convex distribution~\cite{Grover:576789}.
Later extensions~\cite{Cleve_1998,soklakov2004efficient} allow for the preparation of arbitrary states with bounded accuracy with resources scaling polynomially with qubit register size and error.

However, a ubiquitous situation when considering multi-body or discretized approximations of continuous quantum systems is that such states will involve correlations between a large number of single-variable wavefunctions.
Generally, these correlations are fixed by general symmetry principles while the details of the specific state are system dependent.
For example, in the simulation of quantum field theories, bosonic ground states often take the form of a correlated multivariate Gaussian~\cite{Jordan:2011ne}.
For this situation, Kitaev and Webb (KW) suggested an efficient polynomial algorithm consisting of first preparing multiple uncorrelated one-dimensional states and then transforming them to reproduce the required correlations~\cite{kitaev2009wavefunction}.

While the KW algorithm is designed to require polynomial resources in all aspects of multivariate Gaussian preparation, practical considerations lead to a balance of factors when choosing whether asymptotically efficient or \emph{ad hoc} algorithms for finite qubit count should be deployed.
In particular, as observed in~\cite{Macridin:2018gdw, Macridin:2018oli}, if one is interested in low-lying excited states, convergence of simple harmonic oscillator eigenvalues to their continuum values is exponential in qubit count when digitizing in the position (or momentum) basis.
For many field theoretic applications, occupation numbers in any particular mode will never be high, so only a modest number of qubits per continuous degree of freedom may be needed for other sources of error to dominate~\cite{Klco:2018zqz, Bauer:2021gup}.
Meanwhile, extrapolations to the continuum limit are ultimately likely to require a large number of continuous degrees of freedom, so efficient ways of preparing multidimensional ground states will be a necessity.

In light of these considerations, this paper concerned with quantifying the resources required for the implementation of asymptotically efficient multivariate Gaussian-state preparation.
We provide explicit circuits for the various steps in the KW procedure and discuss where these are likely to find use so as to not be the dominant source of error.
For example, errors in the energy eigenvalues of digitized one-dimensional harmonic oscillator states decrease exponentially with the number of state qubits $k$, while we find that circuits using an algorithm polynomially-scaling in $k$ require significant overhead in carrying out the necessary arithmetic.
In particular, in all available implementations, such circuits have a higher entangling gate (CNOT) count than a generic exponentially-scaling algorithm for states with $k \lesssim 15$.
Therefore, simpler exponentially-scaling algorithms are likely to be sufficient for many applications that don't require fine-grained information about the reduced density matrices of one-dimensional subspaces.
In contrast, a polynomial algorithm for the correlation of multidimensional states becomes more efficient for the simulation of all states with more than $\ord(\text{few})$ continuous degrees of freedom.

The remainder of the paper is organized as follows.
In \cref{sec:review}, we review the KW state preparation algorithm, highlighting variations in the implementation that have not previously been clearly stated in the literature.
In \cref{sec:1Dcircuit}, we discuss the resources required for asymptotically efficient preparation of a one-dimensional Gaussian state and compare with the preparation of equivalent states by exponentially-scaling general-state algorithms.
In \cref{sec:shearcircuit} we then present explicit circuits for the rotation of uncorrelated multi-dimensional Gaussian (or in fact any) states into an arbitrary correlated form that is polynomial in gate count, depth, and number of qubits.
We conclude in \cref{sec:conc}.
Many additional details about the algorithms and circuits used are presented in the appendices.

\section{Review of the KW algorithm}
\label{sec:review}

In this section we provide a detailed overview of the algorithm proposed by Kitaev and Webb in~\cite{kitaev2009wavefunction} for producing arbitrary multivariate Gaussian states on a digital quantum computer.
While containing no new results, it provides a pedagogical explanation of the KW procedure, highlights the efficiency provided by quantum parallelism, and presents our notation.
Additionally, we clarify a few points where the KW approach admits variant realizations and discuss their benefits and drawbacks.

\subsection{Discrete representations of Gaussian states}
\label{sec:stateprep}

We start by considering one-dimensional distributions and assuming a scale is chosen such that, when evaluated at integer values, all salient features are captured to the required accuracy.
The Gaussian distribution with mean $\mu$ and variance $\sigma^2$ is defined by
\begin{align}
  \mathcal{N}_{\mu,\sigma}(x) = \frac{1}{\sqrt{2\pi\sigma^2}}\, e^{-\frac{(x-\mu)^2}{2\sigma^2}} \,,
\end{align}
such that a real wavefunction
\begin{align}
  \psi_{\mu,\sigma}(x) = \left(\frac{1}{2\pi\sigma^2}\right)^{1/4} e^{-\frac{(x-\mu)^2}{4\sigma^2}} \,,
\end{align}
will have the property that $|\psi_{\mu,\sigma}(x)|^2 = \mathcal{N}_{\mu,\sigma}(x)$.
The ground state of a simple harmonic oscillator for a particle of mass $m$ centered at $\mu$ corresponds to the choice $\sigma^2 = \hbar/2m\omega$, leading to the ubiquity of the Gaussian as a building block of physical states.

A common approach to representing this wavefunction as the state of some qubits is to discretize the domain of the distribution.
A wavefunction on $n \in \Z$ of the form
\begin{align}
 \label{eq:disc1D}
  \tilde{\psi}_{\mu,\sigma}(n) = \frac{1}{\sqrt{f(\mu,\sigma)}}\, e^{-\frac{(n-\mu)^2}{4\sigma^2}} \,,
\end{align}
is a good approximation for the continuous Gaussian distribution given the earlier assumption about the choice of scale.\footnote{One could alternatively integrate the continuous Gaussian distribution over each desired interval. 
The probabilities of integer values would then be given in terms of error functions. 
Although this would avoid some subtleties involving normalization, the resulting expressions prove to be more complex to numerically approximate than those of the KW algorithm presented here.}
The factor $f(\mu,\sigma)$ appears because the normalization is no longer given by an integral but rather by an infinite sum.
This sum can be evaluated explicitly,
\begin{align}
\label{eq:f}
  \begin{split}
    f(\mu,\sigma) &\equiv \sum_{n=-\infty}^\infty e^{-\frac{(n-\mu)^2}{2\sigma^2}} \\
                  &= \sqrt{2\pi \sigma^2}\, \vartheta\left( \pi\mu; e^{-2\pi^2\sigma^2} \right) \,,
  \end{split}
\end{align}
where $\vartheta(z; \tau)$ is the Jacobi theta function.
It differs from unity by terms exponentially small in the variance,
\begin{equation}
  \label{eq:f_as_theta}
  \vartheta\left( \pi\mu; e^{-2\pi^2\sigma^2} \right)
    = 1 + 2 \sum_{p=1}^\infty \cos(2\pi p\mu)\, e^{-2\pi^2 (p\sigma)^2}  \,,
\end{equation}
such that the normalization is exponentially close to that of the continuous wavefunction for any state with $\sigma^2 \gtrsim 1$.
This allows us to define a quantum state,
\begin{align}
  \label{eq:1Doptstate}
  \ket{\widetilde{\mathcal{N}}_{\mu,\sigma}} = \sum_n \tilde{\psi}_{\mu,\sigma}(n) \ket{n} \,,
\end{align}
such that the probability of measuring $n$ is,
\begin{align}
  \braket{\widetilde{\mathcal{N}}_{\mu,\sigma}}{n} \!\braket{n}{\widetilde{\mathcal{N}}_{\mu,\sigma}}
    \approx \mathcal{N}_{\mu,\sigma}(n) \,,
\end{align}
up to exponentially small correlations.

Ultimately, we are interested in multivariate distributions, for which the above discussion can be straightforwardly generalized.
An $N$-dimensional Gaussian distribution is given by
\begin{align}
\label{eq:multiGauss}
  \mathcal{N}_{\vec{\mu},\Sigma}(\vec{x})
    = \frac{1}{\sqrt{(2\pi)^N \det\Sigma}} \,
        e^{-\frac{1}{2}(\vec{x} - \vec{\mu})^T \Sigma^{-1} (\vec{x} - \vec{\mu})} \,,
\end{align}
with mean vector $\vec{\mu}$ and (real and positive semi-definite) covariance matrix $\Sigma$.
As before, one can define a real wavefunction which satisfies $|\psi_{\vec{\mu},\Sigma}(\vec{x})|^2 = \mathcal{N}_{\vec{\mu},\Sigma}(\vec{x})$.
In analogy to Eq.~\eqref{eq:disc1D}, this can be approximated by a wavefunction defined on a $N$-dimensional integer lattice $\vec{n}$ whose normalization must also be corrected due to discretization.
For this, we do not provide an explicit expression since a different approach to algorithmically approximate the continuous state is pursued below; rather we simply write
\begin{equation}
  \label{eq:NDoptstate}
  \ket{\widetilde{\mathcal{N}}_{\vec{\mu},\Sigma}} =
    \sqrt{C}\, \sum_{\vec{n}} e^{-\frac{1}{4}(\vec{n}-\vec{\mu})^T \Sigma^{-1} (\vec{n}-\vec{\mu})}
      \ket{\vec{n}} \,,
\end{equation}
with normalization constant $C$.
At all lattice sites this state will have expectation values exponentially close to the continuous Gaussian in the same sense as the one-dimensional state above, provided all eigenvalues of $\Sigma$ satisfy $\tilde{\sigma}_i^2 \gtrsim 1$.

In practice, only a finite number of qubits will be available to encode any given state.
Using $k$ qubits, we can encode $2^k$ total points in a given direction, while $Nk$ qubits provide a Hilbert space which can encode a $N$-dimensional lattice of size $2^k$ in each dimension.
If we chose these points to be the finite lattice $B^N_k = [0, \dotsc, 2^k -1]^N$, 
the lattice indices correspond to the computational basis of $N$ $k$-qubit systems.
The whole system can be encoded in the computational basis by concatenating the binary strings for each dimension's coordinates,
\begin{align}
\label{eq:Ndcoords}
  \ket{\vec{n}} \equiv \bigotimes_{i=0}^{N-1} \ket{n_i}, \quad n_i \in B_k \,.
\end{align}
As long as $0 \ll \mu_i \ll 2^k$ and $1 \ll \tilde{\sigma}_i^2 \ll \mu_j^2$ for all means and eigenvalues the truncation (and resulting rescaling) of the wavefunction to this finite lattice will still provide an exponentially accurate approximation to the continuous wavefunction at all lattice sites.

It is this truncation that we will take as the ``optimal'' discretized multivariate Gaussian state when comparing to the results of algorithms discussed in the rest of the paper, leaving detailed questions of how well the resulting state is able to accurately caption the desired continuum dynamics to other work, \eg, \cite{Macridin:2021uwn}.
In particular, since we will remain agnostic about the measurements to be performed on the state after evolution, we adopt $1 - F(\psi,\phi)$ as our measurement of precision, where $F(\psi,\phi) = |\braket{\psi}{\phi}|^2$ is the state fidelity specialized to the case of pure states, the only case of interest in this paper.
We will consider a precise state preparation algorithm one whose state fidelity $F$ is close to 1 when compared with the truncated versions of \cref{eq:1Doptstate,eq:NDoptstate}.

\subsection{Factorization of multivariate Gaussians}
\label{sec:cholesky}

The preceding encoding of a multivariate Gaussian creates a state with a high degree of entanglement between the qubits encoding various dimensions.
However, this entanglement is not arbitrary but fully determined by a covariance matrix with merely $\frac{1}{2}N(N+1)$ independent parameters, which suggests that an efficient way of producing it should exist.

The KW approach accomplishes this by use of the $LDL^T$ decomposition (closely related to the Cholesky decomposition) of the inverse covariance matrix into a diagonal matrix $D$ and a lower unitriangular matrix $L$.
Defining a upper unitriangular matrix $M^{-1} \equiv L^T$, one can write,
\begin{align}
  \Sigma^{-1} = (M^T)^{-1}DM^{-1}
    \quad \Longleftrightarrow \quad
  \Sigma = M D^{-1} M^T \,.
\end{align}
For this choice of parameterization, the coordinate shearing transformation 
\begin{align}
    \vec{y} = M^{-1} (\vec{x} - \vec{\mu})\,,
\end{align} 
lets the exponent in \cref{eq:multiGauss} be factorized into independent terms for each coordinate.
Moreover, $\det M = 1$ and thus $\det \Sigma = \det D^{-1} = \prod_i \sigma_i^2$, where we have defined $\sigma^2_i \equiv D_{ii}^{-1}$.
The associated wavefunction becomes a tensor product of independent normalized, centered Gaussian distributions of varying widths,
\begin{align}
  \psi_{\vec{\mu},\Sigma}(\vec{y})
    = \bigotimes_i \left(\frac{1}{2\pi\sigma_i^2}\right)^{1/4} e^{-y_i^2/4\sigma_i^2}.
    \label{eq:indep_gaussians}
\end{align}

In the (generically not orthonormal) $\vec{y}$ basis, the discretization and preparation of states can be handled individually for each dimension.
For a continuous wavefunction, the inverse shearing transformation, \ie, $\vec{x} = M\vec{y} + \vec{\mu}$, then reproduces the desired state.
Compared to an eigendecomposition of $\Sigma^{-1}$, the $LDL^T$ decomposition has the benefit that the required transformation can be performed \emph{in situ} due to $M$ being upper triangular by successively reordering the indexing of each coordinate, as we discuss in detail below.

For a coordinate lattice, the procedure above can only be implemented approximately.
Discretizing $\vec{y}$ on lattice $\vec{m}$ calls for an approximate inverse coordinate transformation $\vec{n} \approx M \vec{m} + \vec{\mu}$.
If this is accomplished via an overall rounding of the total lattice coordinate, this results in a correction factor per basis state of $\ord(e^{|\vec{n} - \vec{\mu}|)/\tilde{\sigma}_i^2})$, where all quantities are in units of the lattice spacing and by $\tilde{\sigma}_i$ we again indicate some eigenvalue of $\Sigma$.
While this may give large relative corrections for $|\vec{n}-\vec{\mu}| \gg \tilde{\sigma}_i^2$, the amplitudes of such terms are simultaneously exponentially small and absolute corrections are nowhere larger than $\ord(1/\tilde{\sigma}_i)$.
By approximately evaluating the sum of correction factors over all basis states, an overall state fidelity reduction of $\ord(N/\tilde{\sigma}_i)$ is expected when this ratio is small.

The resulting transformation is a mere relabeling of coordinates and can be implemented efficiently on a given coordinate for all lattice sites using quantum parallelism.
This implies that one can divide state preparation into two steps, first preparing independent Gaussian states and then applying a shearing transformation to their coordinates.
For the first step, we discuss an efficient algorithm in \cref{sec:1Drev}, while more aspects of the shearing transformation are discussed in \cref{sec:shearrev}.
Together, they provide a procedure with polynomial scaling in both the number of qubits encoding each continuous degree of freedom $k$ and the number of degrees $N$.

For sufficiently large $k$ and $N$ an exponential scaling in either of these values would be costly.
However, if either parameter is $\ord(\text{few})$, details of implementation will affect the decision to use exponential or polynomial algorithms.
In particular, as we will discuss, applications in field theory often require a large number of degrees of freedom $N$ for reliable extrapolations to continuum results, while relatively modest values of $k$ already lead other aspects of calculations to be the leading sources of error.
As a result, algorithms scaling polynomially with $N$ and exponentially with $k$ can potentially outperform those with uniform polynomial scaling for physically motivated choices of parameters.
%
%Investigating the crossover of these methods is one of our main goals.

\subsection{Efficient 1D state preparation}
\label{sec:1Drev}

For a generic $k$-qubit state, explicit universal state preparation circuits requiring $2^{k+1} - 2k$ entangling CNOT gates for an arbitrary state are known~\cite{Shende_2006}.
For a real wavefunction this can always be reduced to $2^k -2$ CNOTgates, with a symmetric state (in some choice of indexing) admitting a modification using $2^{k-1} + k - 3 + \delta_{1k}$ CNOTs~\cite{Klco:2019xro}. 
Generalizations would allow for a commensurate reduction in gate count for any state displaying $2^n$-fold symmetry, while still retaining a leading $2^{k-n}$ exponential scaling.
Yet, as discussed in the introduction, polynomially scaling algorithms for the preparation of arbitrary states also exist, such that for sufficiently large Gaussian states their use will be preferred.

KW proposed a optimized version of such an algorithm specialized for Gaussian distributions.
To encode a discretized Gaussian on a finite number of qubits, a simple approximation would be to truncate the domain over which the wavefunction is defined.
Instead, the KW approach is to define a periodic function on $k$ qubits by summing over the full discretized Gaussian in windows of the finite lattice $B_k$,
\begin{align}
\label{eq:KWdef}
    \xi^2_{\mu,\sigma;k}(n)
      &= \sum_{m = -\infty}^\infty \tilde{\psi}^2_{\mu,\sigma}(n + m \cdot 2^k) \\
      &= \sum_{m = -\infty}^\infty  \frac{1}{f(\mu,\sigma)}\,
          e^{-\frac{(n + m \cdot 2^k -\mu)^2}{2\sigma^2}} \nnl
      &= \frac{\sqrt{2\pi (\frac{\sigma}{2^k})^2}\,
                 \vartheta\left( \frac{\pi(\mu-n)}{2^k}; e^{-2\pi^2(\sigma/2^k)^2} \right)}
              {f(\mu, \sigma)} \nnl
\label{eq:KWfform}
      &= \frac{f\left( \frac{\mu-n}{2^k}, \frac{\sigma}{2^k} \right)}{f(\mu, \sigma)}\,.
\end{align}
As long as $\mu$ and $\sigma$ take values such that most of the Gaussian is contained in $B_k$, the relative contribution to any basis state from terms in the sum with $m \ne 0$ is exponentially small.
Therefore, the wavefunction
\begin{align}
\label{eq:KWstate}
    \ket{\xi_{\mu,\sigma;k}} \equiv \sum_{n=0}^{2^k-1}\xi_{\mu,\sigma;k}(n)\ket{n} 
\end{align}
has a probability density for $n \in B_k$ which is an exponentially good approximation to a Gaussian distribution.
The advantage of this state is that it admits a recursive definition whose parameters can be given in closed form, as we now prove.

A corollary of \cref{eq:KWdef} is that any sum over $\xi^2_{\mu,\sigma;k}(n)$ in intervals corresponding to powers of 2 can also be written in closed form.
Two simple consequences of the this statement will be useful in their explicit form for the derivation of the recursive definition below.
\begin{description}
  \item[\boldmath $k = 0$]
    The state is simply a sum over the entire $\tilde{\psi}^2(n)$ probability density, so that $\xi_{\mu,\sigma;0}(0) = 1$.
  \item[\boldmath $k = 1$]
    The resulting wavefunction only has 2 components, such that the sum over their probabilities using \cref{eq:KWfform} gives the relation
    \begin{align}
    \label{eq:relation}
      f(\mu, \sigma) &= f\left( \frac{\mu}{2}, \frac{\sigma}{2} \right)
                        + f\left( \frac{\mu-1}{2}, \frac{\sigma}{2} \right).
    \end{align}
  \end{description}
Defining the quantities
\begin{align}
\label{eq:cands}
  c(\mu,\sigma) \equiv \sqrt{\frac{f\left(\frac{\mu}{2},\frac{\sigma}{2}\right)}
                                  {f(\mu,\sigma)}} \, , \quad
  s(\mu,\sigma) \equiv \sqrt{\frac{f\left(\frac{\mu-1}{2},\frac{\sigma}{2}\right)}
                                  {f(\mu,\sigma)}} \,,
\end{align}
both can be written in term of one angle due to \cref{eq:relation},
\begin{align}
\label{eq:alpha}
  \alpha(\mu,\sigma) = \arccos c(\mu,\sigma) = \arcsin s(\mu,\sigma)  \,.
\end{align}

To find the recursive form of $\ket{\xi_{\mu,\sigma;k}}$, we first note that from~\cref{eq:KWdef} it follows that an $\ell$-qubit state observed only for even values of $n$ corresponds to a rescaled $(\ell-1)$-qubit state with a different choice of Gaussian parameters,
\begin{align}
  \xi^2_{\mu,\sigma;\ell}(2n)
    &= \frac{1}{f(\mu,\sigma)} \sum_{m = -\infty}^\infty
         e^{-\frac{(2n + m \cdot 2^\ell -\mu)^2}{2\sigma^2}} \nnl
    &= \frac{1}{f(\mu,\sigma)} \sum_{m = -\infty}^\infty
         e^{-\frac{(n + m \cdot 2^{\ell-1} -\mu/2)^2}{2(\sigma/2)^2}} \nnl
    &= \frac{f\left(\frac{\mu}{2}, \frac{\sigma}{2}\right)}{f(\mu,\sigma)} \, 
         \xi^2_{\frac{\mu}{2},\frac{\sigma}{2};\ell-1}(n) \, .
\end{align}
By a completely analogous calculation an $\ell$-qubit state observed only for odd values of $n$ satisfies
\begin{align}
  \xi^2_{\mu,\sigma;\ell}(2n+1)
    = \frac{f\left(\frac{\mu-1}{2}, \frac{\sigma}{2}\right)}{f(\mu,\sigma)} \,
        \xi^2_{\frac{\mu-1}{2},\frac{\sigma}{2};\ell-1}(n) \,.
\end{align}

\begin{widetext}
The scale factors in these expressions are nothing but the functions defined in \cref{eq:cands}.
The only additional ingredient required is to note that when concatenating strings of qubits to index a given state, the binary identities
\begin{align}
  \ket{2n} = \ket{n} \otimes \ket{0}  \,, \quad
  \ket{2n+1} = \ket{n} \otimes \ket{1}
\end{align}
hold.
Putting these statements together defines any such state recursively in terms of states encoded on one fewer qubit,
\begin{align}
\label{eq:recursive}
  \ket{\xi_{\mu,\sigma;\ell}}
     = \sum_{n=0}^{2^\ell-1} \xi_{\mu,\sigma;\ell}(n) \ket{n}
    &= \sum_{n=0}^{2^{\ell-1}-1} \xi_{\mu,\sigma;\ell}(2n) \ket{2n}
       + \sum_{n=0}^{2^{\ell-1}-1} \xi_{\mu,\sigma;\ell}(2n+1) \ket{2n+1} \nnl
    &= \sum_{n=0}^{2^{\ell-1}-1} c(\mu,\sigma) \,
         \xi_{\frac{\mu}{2},\frac{\sigma}{2};\ell-1}(n) \ket{2n}
         + \sum_{n=0}^{2^{\ell-1}-1} s(\mu,\sigma) \,
           \xi_{\frac{\mu-1}{2},\frac{\sigma}{2};\ell-1}(n) \ket{2n+1} \nnl
%    &= \sum_{n=0}^{2^{\ell-1}-1} \xi_{\frac{\mu}{2},\frac{\sigma}{2};\ell-1}(n) \ket{n}
%         \otimes c(\mu,\sigma) \ket{0}
%         + \sum_{n=0}^{2^{\ell-1}-1} \xi_{\frac{\mu-1}{2},\frac{\sigma}{2};\ell-1}(n) \ket{n}
%             \otimes s(\mu,\sigma) \ket{1} \nnl
    &= \ket{\xi_{\frac{\mu}{2},\frac{\sigma}{2};\ell-1}} \otimes \cos \alpha(\mu,\sigma) \ket{0}
       + \ket{\xi_{\frac{\mu-1}{2},\frac{\sigma}{2};\ell-1}} \otimes \sin \alpha(\mu,\sigma) \ket{1} \,.
\end{align}
Explicit examples of the first few such recursively-constructed states are provided in \cref{app:recursionex}.
\end{widetext}

This recursive definition was used by KW to provide a method for the preparation of any $\ket{\xi_{\mu,\sigma;k}}$.
The required state is constructed recursively by first rotating the final qubit by the angle $\alpha(\mu,\sigma)$, and then building up states with lower $\ell$ through unitaries controlled by already prepared qubits.
Denoting the depth of recursion by $j$, the values of $\mu_j$ and $\sigma_j$ (and hence $\alpha_j \equiv \alpha(\mu_j, \sigma_j)$) required for the full construction of the state are
\begin{align}
\label{eq:mu_j} 
  \sigma_0 &= \sigma
    & \mu_0 &= \mu \\
  \sigma_{j+1} &= \frac{1}{2}\sigma_j
    & \mu_{j+1} &= \begin{cases}
                     \frac{\mu_j}{2}   \quad &\text{if } \ket{q_j} = \ket{0} \\
                     \frac{\mu_j-1}{2} \quad &\text{if } \ket{q_j} = \ket{1}
                   \end{cases} \,,
\end{align}
where $\ket{q_j}$ denotes the qubit rotated by $\alpha_j$.
With this notation, the states that need to be prepared at the $j^\text{th}$ iteration are $\ket{\xi_{\mu_j,\sigma_j;k-j}}$ defined by \cref{eq:recursive} corresponding to the $2^j$ different values $\mu_j$ takes.
This procedure terminates when $j = k$ with a final rotation since $\ket{\xi_{\mu_k,\sigma_k;0}} = 1$. 
A high-level description of such an algorithm is

\begin{algorithm}[H]
  \DontPrintSemicolon
  \KwData{$\{\mu, \sigma\}$, state size $k$, angle precision $b$}
  \KwResult{state $\ket{\xi_{\mu,\sigma;k}} \otimes \ket{0}^{\otimes b}$}
%            \otimes \ket{0}^{\otimes \norm{\text{anc}}}$}
  \Begin{
    $\sigma_0 \leftarrow \sigma$\;
    state register $\ket{q_{k-1} \dotsi q_0} \gets \ket{0}^{\otimes k}$\;
    angle register $\ket{\alpha} \gets \ket{0}^{\otimes b}$\;
%    arithmetic ancilla register $\ket{a} \gets \ket{0}^{\otimes \norm{\text{anc}}}$\;
    \For{$j \gets 0$ \KwTo $k-1$}{
      given $\mu$, interpret $\ket{q_{j-1} \dotsi q_0}$ as $\ket{\mu_j}$\;
      $\ket{\mu_j} \ket{\alpha=0} \gets \ket{\mu_j} \ket{\alpha_j = \alpha(\mu_j,\sigma_j)}$\;
      $\ket{q_j = 0} \ket{\alpha_j} \gets
       \left( \cos \alpha_j \ket{0} + \sin \alpha_j \ket{1} \right) \ket{\alpha_j}$ \;
      uncompute $\alpha_j$: $\ket{\mu_j} \ket{\alpha_j} \gets \ket{\mu_j} \ket{0}^{\otimes b}$\;
      $\sigma_{j+1} \gets \sigma_j/2$\;
    }
  }
\caption{1D Gaussian state preparation}
\label{alg:1Dprep}
\end{algorithm}

If all parameters are computed classically, the resulting algorithm requires exponential resources.
Each additional state qubit doubles the number of rotations at the last recursive step, each with a different value of $\mu_j$ (and hence $\alpha_j$), leading to a total of $2^k-1$ separate angle evaluations~\cite{Klco:2019xro}.
However, the KW procedure takes advantage of the fact that the calculations required for each rotation are identical up to inputs conditioned on state qubits that have already been set.
All rotations on a given qubit can then be performed on a superposition of states, and as long as all arithmetic uses polynomial resources, the entire algorithm has polynomial scaling.

Given that arithmetic to compute the angle $\alpha_j$ to $b$-qubit accuracy on a quantum computer (as on a classical computer) can be implemented with reversible gates and polynomial scaling, this algorithm is efficient.
This leads to (sub)linear scaling in $k$ since despite the doubling of recursive states in \cref{eq:recursive}, all states for each $j$ are prepared in parallel using a single arithmetic circuit to compute $\alpha(\mu_j,\sigma_j)$.
When applied to qubits storing a superposition of $\ket{\mu_j}$ values, it computes the corresponding superposition $\ket{\alpha(\mu_j,\sigma_j)}$.
Applying a $R_y(2\alpha_j)$ rotation to $\ket{q_j}$ then prepares the $j^\text{th}$ state qubit in the appropriately entangled superposition with the already computed $\ket{q_{j-1} \dotsi q_0}$.

In contrast to the original KW presentation, we note the lack of any ancillae to store parameters used in the computation of the angles $\alpha_j$.
The variances $\sigma_j^2$ can be computed classically since they take a unique value for each $j$, while the means $\mu_j$ do not require a dedicated register since their value can be efficiently extracted from the already-computed state qubits.
In \cref{sec:1Dcircuit}, we explain further the preceding claims and derive resource costs for explicit quantum circuits implementing this algorithm.

\subsection{Shearing transformation}
\label{sec:shearrev}

The shearing transformation acts on the integer-valued state coordinates $\vec{m}$ to transform them into another set of integer-valued vectors $\vec{n}$ according to the transformation $\vec{n} \approx M \vec{m} + \vec{\mu}$.
Here we write $\approx$ since the requirement that the output remain integer-valued requires some form of rounding.
At the level of the state, this corresponds to some permutation of amplitudes in the computational basis.
In the following, we will assume that $ \vec{\mu} = 0$, but the discussion is easily generalized for arbitrary Gaussian mean.
Note that as the matrix $M$ is upper unitriangular, this transformation can be written as
\begin{align}
\label{eq:Mrel}
  n_i = m_i + R\left[\sum_{j=i+1}^N M_{ij} m_j \right] \,,
\end{align}
such that the values $n_i$ only depend on $m_{j \geq i}$.
Each $m_i$ can be transformed into $n_i$ by a permutation in place, with the resulting transformation of the physical state not affecting the shearing of coordinates with higher $i$.

In general, $M_{ij}$, and hence $\sum_{j=i+1}^N M_{ij} m_j$, is not integer valued.
Since the $n_i$ on the left-hand side of \cref{eq:Mrel} are required to be integers, this equation requires a rounding prescription, indicated by the function $R[\,\cdot\,]$.
Various choices for the rounding function are possible, for example one can first perform the sum and then round, or round each element in the sum.
The latter was advocated in the original KW algorithm, while we implement --- and recommend --- the former.

Another ambiguity is that the resulting value $n_i$ is not guaranteed to be in the range $B_k$.
This implies that a prescription needs to be defined how to deal with such overflow values.
An option we adopt requiring no garbage collection, which would be costly to implement, is to perform all lattice coordinate transformations with modulo $2^k$ arithmetic.
It should be noted that for Gaussian states centered at $n_i \approx 2^{k-1}$, the amplitudes of lattice sites near the edge of the lattice are exponentially suppressed, and a correspondingly exponentially small effect due to the choice of prescription will be observed.

Finally, one needs to remember that the Hilbert space spanning $\ket{\vec{n}}$ has dimension $2^{Nk}$, so that a classical computation of the shearing operation given in \cref{eq:Mrel} requires exponential resources both in the number of qubits per lattice site as well as the number of lattice sites.
A state preparation algorithm with polynomial scaling requires implementing the shearing transformation again using quantum parallelism.
Explicit circuits implementing the shearing transformation and a further discussion of the variants in implementation mentioned above are presented in \cref{sec:shearcircuit}.

\section{1D Gaussian preparation}
\label{sec:1Dcircuit}

\Cref{sec:1Drev} highlighted which parts of Gaussian state preparation in the KW procedure require a quantum calculation in order to achieve scaling polynomial in the number of state qubits $k$ and angle qubits $b$, but did not present an explicit realization.
An outline and accounting of the resource requirements of such a construction is the goal of this section.
Minimizing the resource requirements requires considering the optimization of quantum arithmetic circuits and to this purpose we provide a short review of the existing literature on the subject in \cref{app:qmathsummary}.

It is important to keep in mind that when the final goal is the construction of some multidimensional Gaussian state, the first part of the KW algorithm by construction produces no entanglement between the different dimensions.
Thus, instead of having the complexity of a general operation on the full Hilbert space of dimension $2^{Nk}$, the effective dimensionality of an arbitrary initial state before applying a shearing transformation is only $N 2^k$, and already scales linearly with the number of dimensions.
Ultimately, we will find that for small values of $k$, as is typically sufficient to avoid being the leading source of error for many applications, the KW algorithm with polynomial scaling in $k$ is much more costly, both in gate count and ancillae, than a generic exponentially scaling algorithm for real symmetric state preparation in each dimension.

\subsection{\boldmath Required qubit registers}
\label{sec:registers}

Explicit circuits to implement every step in the recursive procedure of \cref{alg:1Dprep} require the qubit registers
\begin{itemize}
  \setlength{\itemsep}{2pt}
  \item $\ket{q} \equiv \ket{q_{k-1} \dotsi q_0}$ for the state $\ket{\xi_{\mu, \sigma; k}}$,
  \item $\ket{\tilde{\alpha}}$ for a $b$-qubit representation of $\alpha$,
\end{itemize}
along with a number of ancilla registers to store intermediate results of the angle calculation, which are discussed in more detail in \cref{app:computealpha}.
We denote the angle here by $\tilde{\alpha}$ to indicate that the value stored in the register is actually $2\alpha/\pi$ such that all possible rotations are stored as a binary fraction between 0 and 1.
The angle and ancilla registers can be reused between different 1D Gaussian states, so the number of work qubits does not grow with the number of dimensions.

While the various rotation angles used by the recursive algorithm are computed in superpostion, there is no need to store their associated state parameters in dedicated registers.
Since $\sigma_j = \sigma/2^j$ takes a unique value at each recursion level it can be incorporated by adjusting the various constants required by the circuit (see \cref{app:approxalpha} for details), while $\mu_j$ is already implicitly stored in the state register in a form suitable for calculation.

Recall the recursive definition of $\mu_{j}$,
\begin{align}
\label{eq:mu_expression}
  \mu_{j} = \begin{cases}
                \frac{\mu_{j-1}}{2}     &\text{if } \ket{q_{j-1}} = \ket{0} \\
                \frac{\mu_{j-1} - 1}{2} &\text{if } \ket{q_{j-1}} = \ket{1}
              \end{cases}.
\end{align}
At the $j^\text{th}$ iteration, $\mu_j$ takes $2^j$ possible values, and the bifurcation from the previous iteration is determined by the state of $\ket{q_{j-1}}$. 
But this means that $\ket{q_{j-1} \dotsi q_0}$ uniquely determines the bifrucation history, and thus the given value of $\mu_j$ to be used in computing $\alpha(\mu_j,\sigma_j)$.
Moreover, as this mapping is linear there is no benefit to storing $\ket{\mu_j}$ in a dedicated register rather than simply interpreting the value of the subregister $\ket{q_{j-1} \dotsi q_0}$ as $\ket{\mu_j}$ itself.

To be more explicit, given a computed state $\ket{q_{j-1} \dotsi q_0}$ the values of $\mu_j$ appearing at recursion level $j$ are
\begin{align}
\label{eq:muj_explicit}
  \mu_j = \frac{\mu - (q_{j-1} \dotsi q_0)_2}{2^j},
\end{align}
where $({}\dotsi)_2$ indicates the binary representation of some integer.
This takes on a particularly simple form when $\mu = -\frac{1}{2}$.
In that case, writing $\mu_j$ in binary notation,
\begin{align}
\label{eq:mu5_explicit}
  \mu_j = -(0.q_{j-1} \dotsi q_0 1)_2,
\end{align}
and the state of $\ket{q_{j-1} \dotsi q_0}$ can be directly taken as a superposition of the required binary expansions of $\mu_j$.
It is immediately clear that in this case $\mu_j \in (-1, 0)$ at every recursion level.\footnote{In fact, this it true for any $\mu \in [-1,0]$, so that all possible Gaussian states, including those with no particular discretized symmetry properties, can be prepared in a similar manner.}
Being constant, both the sign and the least significant bit can simply be appended when required for calculation at low implementation cost.

The decomposition of \cref{sec:cholesky} ensures that when preparing the initial unentangled 1D states, we are free to set their means to whatever value is most convenient for implementation, since it can always be compensated by an overall shift when later applying the shearing transformation.
Since simplifications occur if we can interpret the state as being symmetrically encoded on some choice of qubits, the considerations above motivate the reinterpretation of our fundamental finite lattice as defined on the sites $B'_k = [-2^{k-1}, \dotsc, 2^{k-1} -1]$, leading to a symmetric encoding of the $\mu = -\frac{1}{2}$ state.
From a computational perspective this is nothing but the reinterpretation of the computational basis in two's complement to encode signed lattice coordinates.
As the states $\ket{\xi_{\mu,\sigma;k}}$ are periodic by construction, this redefinition of the primary lattice leaves all prior expressions unmodified.

\subsection{\boldmath Computing \texorpdfstring{$\ket{\alpha_j}$}{the rotation angle}}
\label{sec:computealpha}

All data required for computing the required superposition of rotation angles are already encoded in the state qubits, but we are still left with the task of implementing the necessary arithmetic.
This is not a straightforward task, as the evaluation of $\alpha(\mu_j,\sigma_j)$ as defined by \cref{eq:alpha} requires the computation of a number of functions whose direct implementation on quantum hardware has not received much study to our knowledge.
To construct an efficient implementation, we have chosen to implement a combination of three approximations which can be classically selected between at each recursion level using the value of $\sigma_j$.
As the resulting circuits are fairly involved, we reserve their explicit presentation to \cref{app:polycirc,app:piececirc}.
Moreover, we find that the resource requirements are such that for small per-dimension registers, exponentially-scaling algorithms with lower overhead are preferable.
However, we briefly discuss the functional form of $\alpha(\mu_j,\sigma_j)$ at fixed $\sigma_j$ here to provide a high-level overview of our implementation.

As discussed in \cref{sec:registers}, for a judicious choice of indexing, we can always ensure $\mu_j \in (-1, 0)$.
Therefore, we only need quantum circuits capable of evaluating $\alpha_j$ constrained to the range $\mu_j \in (-1,0)$ for fixed values of $\sigma_j$.
As a function of $\mu_j$, $\alpha(\mu_j,\sigma_j)$ is always a function with period 2 with no simple rapidly converging universal approximation.
However, at very large and very small $\sigma_j$, it can be well approximated by simpler functions.

For large $\sigma_j$, \cref{eq:alpha,eq:f_as_theta} lead to an expansion of $\alpha(\mu_j,\sigma_j)$ in simple trigonometric functions of $\mu_j$ around an average value of $\pi/4$ whose coefficients are exponentially suppressed by $\sigma_j^2$.
In this case we can simply evaluate the Taylor series approximation of the resulting function to the required accuracy.
For small $\sigma_j$, $\alpha(\mu_j,\sigma_j)$ is exponentially close to a square wave of amplitude $\pi/2$, which has an even simpler implementation as a single controlled rotation.
For intermediate regions with $\sigma_j \sim \ord(1)$, the general functional behavior is complex, with periodic variation growing as $\sigma_j$ decreases until it saturates to the above square wave behavior.
In this region we implement a piecewise series approximation for different regions of $\mu_j$, evaluated in parallel using the methods of \cite{haner2018optimizing}.

In all cases, the angle register size determines the accuracy to which $\alpha_j$ should be computed, and thus selects the order of approximation used in the arithmetic circuits to be such that all errors are kept smaller than the resolution of the angle register at each step.
All resource costs and state fidelities below are thus quoted truncating approximations at the optimal order for a given register size.

\begin{figure}[bp]
  \begin{adjustbox}{width=\columnwidth}
    \begin{tikzcd}
       \lstick{$\ket{\tilde{\alpha}_j}_{b-1}$}
         & \ctrl{4} & \qw      & \qw & \ \ldots\ \qw & \qw & \qw       & \qw \\
       \lstick{$\ket{\tilde{\alpha}_j}_{b-2}$}
         & \qw      & \ctrl{3} & \qw & \ \ldots\ \qw & \qw & \qw       & \qw \\
       \vdots \hspace{40pt} &&&& \vdots &&&&& \\
       \lstick{$\ket{\tilde{\alpha}_j}_{0\phantom{-1}}$}
         & \qw      & \qw      & \qw & \ \ldots\ \qw & \qw & \ctrl{1}  & \qw \\
       \lstick{$\ket{q_j}$}
         & \gate{R_y\left(\frac{\pi}{2}\right)}
           & \gate{R_y\left(\frac{\pi}{4}\right)} & \qw & \ \ldots\ \qw
           & \qw & \gate{R_y\left(\frac{\pi}{2^b}\right)} & \qw
    \end{tikzcd}
 \end{adjustbox}
 \caption{Circuit implementing $R_y(2\alpha_j)$ on $\ket{q_j}$.}
 \label{fig:qc_rotate_alpha}
\end{figure}
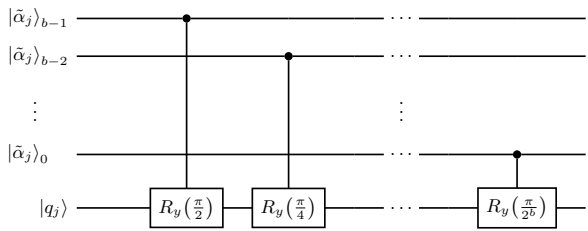

\subsection{\boldmath Rotating \texorpdfstring{$\ket{q_j}$}{the next state qubit}}

Once $\tilde{\alpha}_j = 2\alpha_j/\pi$ has been computed, the rotation $R_y(2\alpha_j)$ needs to be applied to $\ket{q_j}$.
Since $\tilde{\alpha}_j$ is already stored as a $b$-qubit binary fraction, it can be written as
\begin{align}
    \alpha_j = \sum_{r=0}^{b-1} \frac{\pi}{2} \frac{\tilde{\alpha}_{j,r}}{2^{b-r}} \,,
\end{align}
where $\tilde{\alpha}_{j,r}$ indicates the value of the $r$th qubit encoding $\tilde{\alpha}_j$.
With all $R_y$ rotations commuting, $R_y(2\alpha)$ can be broken into $b$ individual rotations,
\begin{align}
    R_y(2\alpha_j)
      = \prod_{r=0}^{b-1} R_y\left(\frac{\pi}{2^{b-r}} \tilde{\alpha}_{j,r} \right) \,,
\end{align}
and implemented as a product of fixed angle rotations controlled by $\ket{\tilde{\alpha}_j}_r$, as illustrated in \cref{fig:qc_rotate_alpha}.
In other words, $R_y(2\alpha)$ can be implemented as a sequence of $b$ controlled rotations, and thus requires only $2b$ CNOTS using standard controlled unitary decompositions~\cite{Barenco:1995na}.

There are two further simplifications worth remarking on, which, as with the angle computation circuits above, are addressed in more detail in \cref{app:computealpha}.
For low values of $\sigma_j$ when the angle can be well-approximated by a square wave in $\mu_j$, the angle calculation and state qubit rotation can be combined into one step requiring only a single CNOT with no uncomputation.
Similarly, for any angle register, there will be values of $\sigma_j$ above which $\alpha_j \approx \pi/4$ within the resolution of the $b$-qubit angle register.
In this case, no actual $\mu_j$-dependent calculation needs to be performed since the state qubit rotation can be implemented by an uncontrolled $R_y(\pi/2)$ (or $H$ since it always acts on an unmodified $\ket{q_j = 0}$) gate.
Thus both very large and very small values of $\sigma_j$ will use few resources, and the dominant resource costs occur at recursion levels for which $\sigma_j$ takes intermediate values.

\begin{figure}[tbp]
  \subfloat[Fixed $\sigma$]{
    \includegraphics[width=0.8\columnwidth]{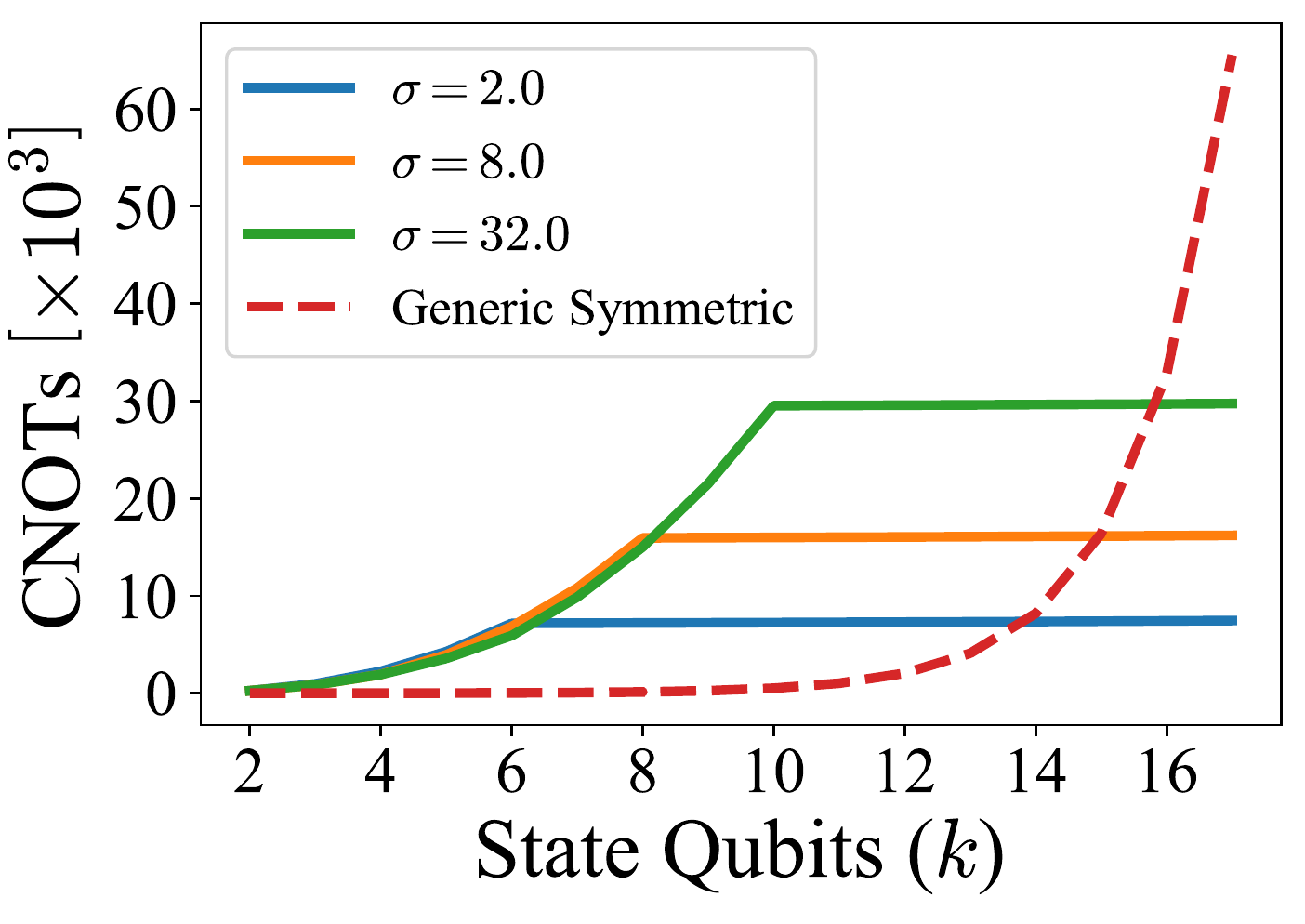}
    \label{fig:cnotfix}
  } 
  
  \subfloat[Physical $\sigma$]{
    \includegraphics[width=0.8\columnwidth]{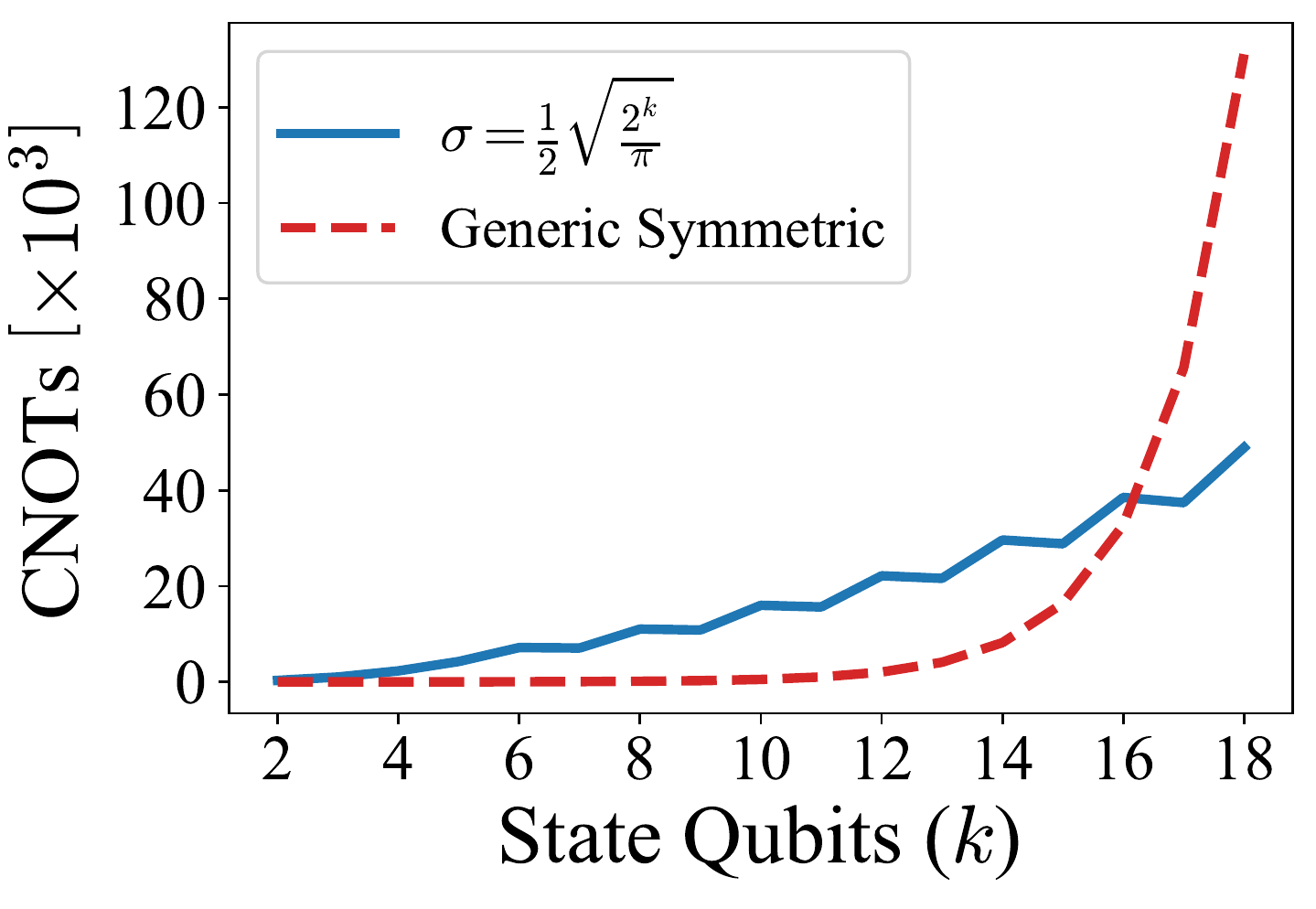} 
    \label{fig:cnotphys}
  }
  
  \caption{Entangling gate cost comparison for 1D Gaussian state preparation methods as a function of state register size. Exponential refers to the method optimized for symmetric real wavefunctions of \cite{Klco:2019xro} requiring no work qubits but scaling exponentially with number of state qubits. The polynomially-scaling KW method is shown in two scenarios. Several fixed choices of state width are plotted in (a), while field digitization optimized for simple harmonic oscillator eigenvalue accuracy~\cite{Bauer:2021gup} is shown in (b).}
  \label{fig:cnots_1dprep}
\end{figure}

\subsection{Resource requirements}

Having outlined the various circuits required to implement the 1D Gaussian preparation procedure of KW, we quantify their costs and performance.
\Cref{fig:cnots_1dprep} illustrates CNOT gate costs for our implementation of \cref{alg:1Dprep}.
This is compared against the method of~\cite{Klco:2019xro}, optimized for the preparation of symmetric real wavefunctions, but employing no quantum parallelism and resulting in an overall exponential scaling.

We show the resulting gate costs for two different scenarios.
In \cref{fig:cnotfix} the preparation costs of various Gaussians of fixed width are displayed, while \cref{fig:cnotphys} uses a width scaling as a function of $k$-qubit register size, advocated in~\cite{Bauer:2021gup} as optimal for the Hamiltonian of a simple harmonic oscillator.
In the latter case, $\sigma \propto 2^{k/2}$, and doubles with every 2 additional qubits.
Therefore, for any choice of thresholds for the approximations described in \cref{sec:computealpha}, one extra computation in the large $\sigma_j$ regime is required for every two additional qubits in the state register, resulting in the step-like behavior of \cref{fig:cnotphys}.
When keeping width fixed, the marginal gate costs become negligible after $\sigma_j$ becomes low enough that the square-wave approximation of the rotation angle becomes sufficient.

\begin{figure}[tp]
    \includegraphics[width=0.9\columnwidth]{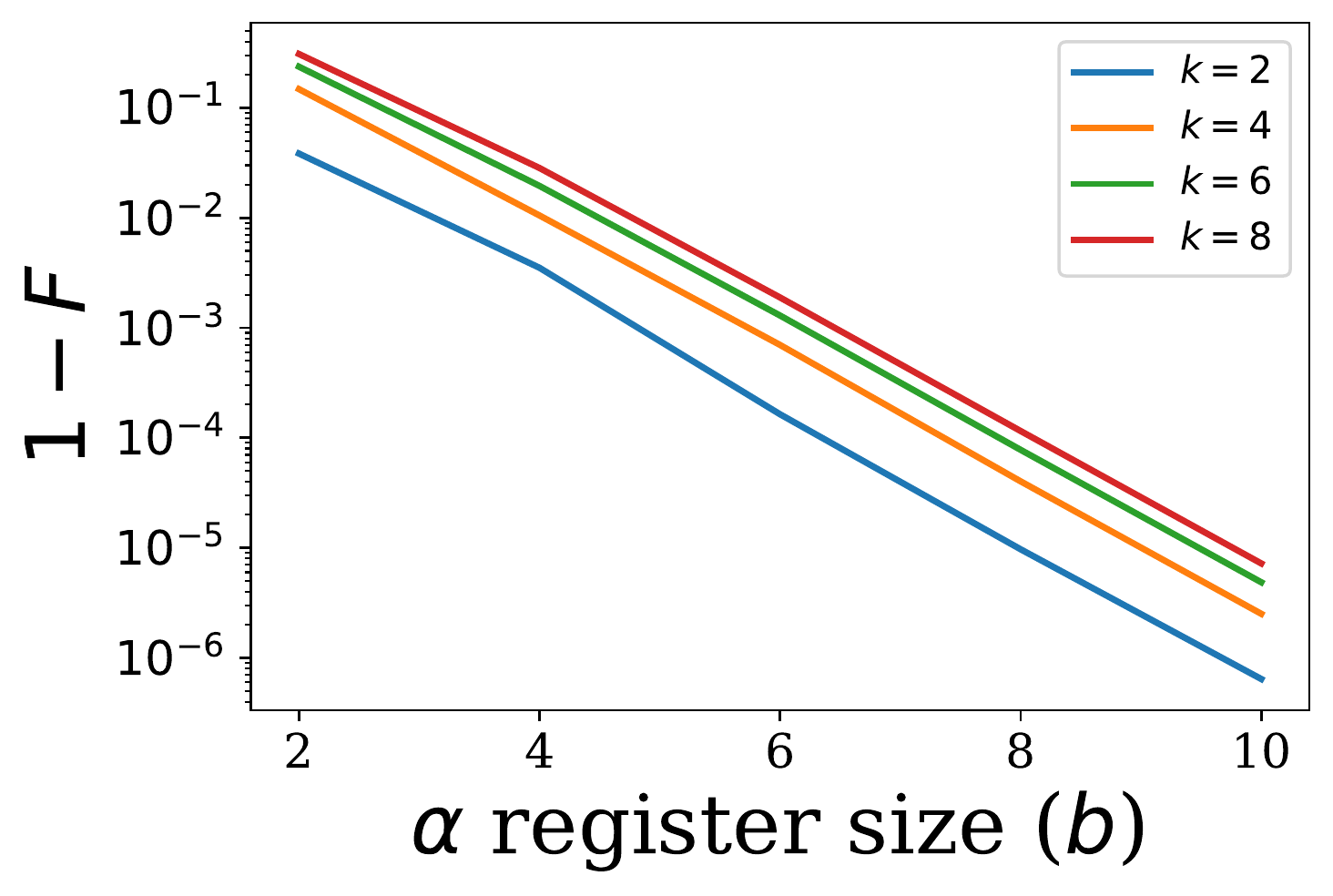}
    \caption{Pure state fidelity $F = |\braket{\psi}{\phi}|^2$ between the exact KW state $\ket{\xi_{\mu,\sigma;k}}$ and one created using a polynomially-scaling algorithm using a $b$-qubit register for angle $\alpha_j$ in all arithmetic circuits.}
    \label{fig:fidelity_alphaqubits}
\end{figure}

\Cref{fig:fidelity_alphaqubits} illustrates the effect of a finite $\alpha(\mu_j,\sigma_j)$ register size on the accuracy of the KW 1D Gaussian algorithm.
The fidelity between the exact state $\ket{\xi_{\mu,\sigma;k}}$ and a state prepared using angles stored in an angle register of $b$-qubit size is shown as a function of state register size.
The exponential improvement in state fidelity with increasing angle register size is clear, while overall fidelity for a given choice of $b$ has a weak negative dependence on the size of the state to be prepared.

\section{\boldmath \texorpdfstring{$N$}{N}-dimensional shearing}
\label{sec:shearcircuit}

As described in \cref{sec:shearrev}, the shearing coordinate transformation needs to implement the mapping \cref{eq:Mrel} on the state indices of \cref{eq:Ndcoords}.
The resulting state bases become
\begin{align}
\label{eq:quantum_shearing_trafo}
    \ket{n_i} = \ket{m_i + R\left[\sum_{j=i+1}^{N-1} M_{ij}m _j\right]} \,,
\end{align}
with the rounding operation $R[\,\cdot\,]$ accounting for the fact that $n_i$ needs to be an integer.
Furthermore, all arithmetic is applied modulo $2^k$, such that the transformation on each coordinate's $k$-qubit register corresponds to some unitary permutation of basis states.
The notation above assumes that the Gaussian is centered at the origin of the lattice, and in cases where this is not the case, the appropriate offsets need to be included in the formulas and circuits.

A general state on $\ket{\vec{n}}$ is defined in a Hilbert space of dimension $2^{Nk}$, and the transformation given by \cref{eq:quantum_shearing_trafo} is potentially different for each of these states. 
A classical computation to perform an arbitrary permutation on this basis would require resources exponential in $N$ and $k$, becoming prohibitive to implement for high-dimensional states.
A quantum arithmetic circuit implementing \cref{eq:quantum_shearing_trafo} on a superposition of different $\ket{\vec{n}}$ can transform an entire dimension simultaneously, however.
For this to be possible without large ancilla and garbage collection costs, it is crucial that $\ket{n_i}$ only depends on $\ket{m_j}$ with $j \geq i$. 
Thus the $\ket{n_i}$ can be computed in place --- meaning $\ket{m_i}$ is mapped to $\ket{n_i}$ --- beginning with $\ket{m_0} \gets \ket{n_0}$ and proceeding to $\ket{m_{N-2}} \gets \ket{n_{N-2}}$. 
(The mapping $\ket{m_{N-1}} \gets \ket{n_{N-1}}$ is always an identity and requires no calculation.)
This feature is essential for an efficient quantum circuit implementation.

The shearing matrix $M$ is a classical entity determined by the desired multivariate Gaussian state. 
This allows the product $M_{ij} m_j$ to be efficiently implemented on a quantum circuit using a hybrid quantum-classical version of the standard shift-and-add (S/A) multiplication algorithm. 
The product of $M_{ij}$ and a quantum register $\ket{m_j}$ is computed by looping over the binary representation of $M_{ij}$ and for each 1 adding $m_j$ with the appropriate binary offset to the product register.

Compared to existing quantum multiplication circuits based on the S/A algorithm~\cite{munozcoreas2017_mult}, classically-controlled multiplication (CCM) uses significantly fewer gates.
First, CCM uses fixed adders, as opposed to the controlled adders used in S/A multiplication, which are more expensive to implement reversibly.
Second, the number of adders required to compute $M_{ij}m_j$ is equal to the number of 1s in the binary representation of $M_{ij}$, whereas the number of controlled adders used in a standard quantum multiplication circuit is equal to the total length of bitstring $M_{ij}$.
As CCM encodes the value of $M_{ij}$ directly into the circuit, only 2 quantum registers, the multiplicand $\ket{m_j}$ and the product register, are required.
As with typical S/A multiplication, if the product register is initialized to some non-zero value, the product is added to this value, and successive applications of CCM to a coordinate register will implement the summation of terms in \cref{eq:quantum_shearing_trafo}.
We present more details and an explicit circuit realization of CCM itself in~\cref{app:CCMA}.\footnote{If the product register is known to be set to 0 initially, the CNOT count can be reduced by eliminating many of the carry operations, since each partial sum is guaranteed to not generate carries beyond its highest qubit.}
Applying CCM whenever possible, an efficient algorithm for implementing the shearing transform is
\begin{algorithm}[H]
  \DontPrintSemicolon
  \KwData{$M_{ij}$, $\psi(\vec{m})\ket{\vec{m}}$}
  \KwResult{state $\psi(\vec{m})\ket{\vec{n}} \otimes \ket{0}^{\otimes 2r}$ via \cref{eq:quantum_shearing_trafo}}
  \Begin{
    $\ket{e}, \ket{f} \gets \ket{0}^{\otimes r}$\;
    \For{$i \gets 0$ \KwTo $N-2$}{
      $\ket{\tilde{n}_i} = \ket{m_i.0 \dotsm} = \ket{m_i} \otimes \ket{f}$\;
      \For{$j \gets i+1$ \KwTo $N-1$}{
        $\ket{\tilde{m}_j} = \ket{e} \otimes \ket{m_j}$ \;
        $\ket{\tilde{m}_j} \gets$ sign extension (SE) of $\ket{m_j}$ \;
        $\ket{\tilde{n}_i} \gets \ket{\tilde{n}_i} + M_{ij} \ket{\tilde{m}_j}$
          \tcp*[r]{via CCM}
        $\ket{\tilde{m}_j} \gets$ SE$^{-1}$ of $\ket{m_j}$ \;
      }
      \uIf(\tcp*[f]{for rounding}){
        $\ket{f} > \frac{1}{2}$}{$\ket{\tilde{n}_i} \gets \ket{\tilde{n}_i} + 1$ \;
      } 
      \For{$j \gets N-1$ \KwTo $i+1$}{
        $\ket{\tilde{m}_j} = \ket{e} \otimes \ket{m_j}$ \;
        $\ket{\tilde{m}_j} \gets$ SE of $\ket{m_j}$ \;
        $\ket{f} \gets \ket{f} - M_{ij} \ket{\tilde{m}_j}$
          \tcp*[r]{via CCM$^{-1}$}
        $\ket{\tilde{m}_j} \gets$ SE$^{-1}$ of $\ket{m_j}$ \;
      }
    }
  }
  \caption{Coordinate shearing transform}
  \label{alg:shear}
\end{algorithm}

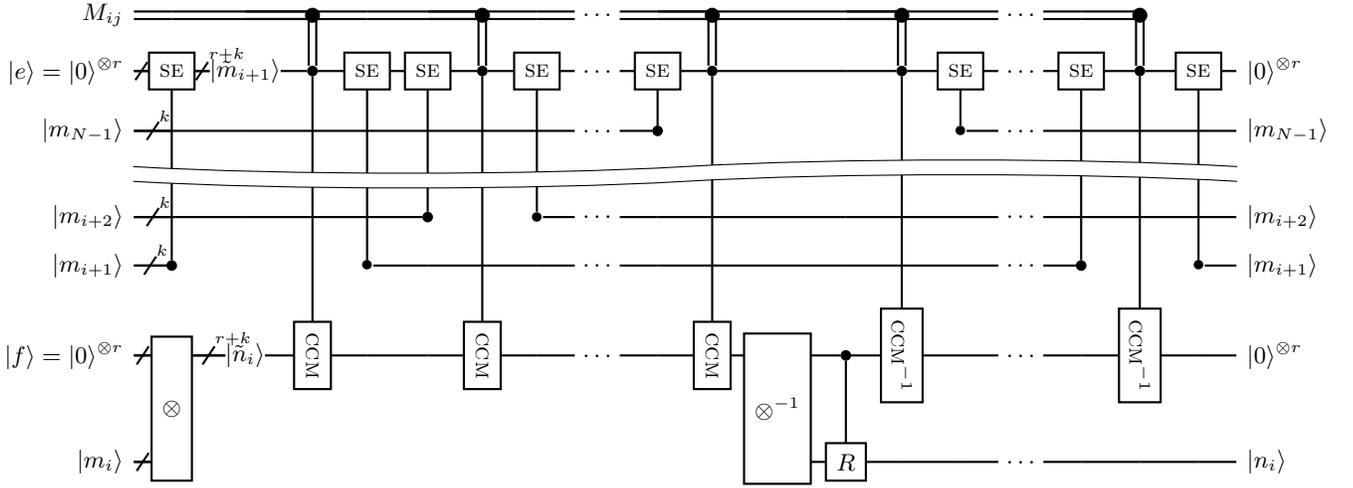
\begin{figure*}
  \centering
  \begin{adjustbox}{width=\textwidth}
  \begin{tikzcd}[column sep = 0.2 cm]
    \lstick{$M_{ij}$} & \cw & \cw & \cwbend{1}\cw & \cw & \cw & \cwbend{1}\cw & \cw & \ \ldots\ \cw & \cw & \cwbend{1}\cw & \cw & \cw & \cwbend{1}\cw & \cw & \ \ldots\ \cw & \cw & \cwbend{1} &&\\
    \lstick{$\ket{e} = \ket{0}^{\otimes r}$} & \gate{\textsc{se}}\qwbundle{r}
      & \push{\ket{\tilde{m}_{i+1}}}\qwbundle{r+k} & \ctrl{5}  & \gate{\textsc{se}}
      & \gate{\textsc{se}} & \ctrl{5} & \gate{\textsc{se}} & \ \ldots\ \qw & \gate{\textsc{se}} & \ctrl{5} & \qw & \qw & \ctrl{5} & \gate{\textsc{se}} & \ \ldots\ \qw & \gate{\textsc{se}} & \ctrl{5} & \gate{\textsc{se}} & \rstick{$\ket{0}^{\otimes r}$}\qw\\
    \lstick{$\ket{m_{N-1}}$} & \qw\qwbundle{k} & \qw & \qw & \qw & \qw & \qw & \qw
      & \ \ldots\ \qw & \ctrl{-1} & &&&& \bullet\vqw{-1} & \ \ldots\ \qw & \qw & \qw & \qw
      & \rstick{$\ket{m_{N-1}}$}\qw\\
    \wave &&&&&&&&&&&&&&&&&&&& \\
    \lstick{$\ket{m_{i+2}}$} & \qw\qwbundle{k} & \qw & \qw & \qw & \ctrl{-3} && \bullet\vqw{-3}
      & \ \ldots\ \qw & \qw & \qw & \qw & \qw & \qw & \qw & \ \ldots\ \qw & \qw & \qw & \qw
      & \rstick{$\ket{m_{i+2}}$}\qw\\
    \lstick{$\ket{m_{i+1}}$} & \ctrl{-4}\qwbundle{k} &&& \bullet\vqw{-4} & \qw & \qw & \qw
      & \ \ldots\ \qw & \qw & \qw & \qw & \qw & \qw & \qw & \ \ldots\ \qw & \ctrl{-4}
      && \bullet\vqw{-4} & \rstick{$\ket{m_{i+1}}$}\qw \\
    \lstick{$\ket{f} = \ket{0}^{\otimes r}$} & \gate[wires=2]{\otimes}\qwbundle{r}
      & \push{\ket{\tilde{n}_i}}\qwbundle{r+k} & \gate{\rotatebox{270}{\textsc{ccm}}} & \qw & \qw
      & \gate{\rotatebox{270}{\textsc{ccm}}} & \qw & \ \ldots\ \qw & \qw
      & \gate{\rotatebox{270}{\textsc{ccm}}} & \gate[wires=2, nwires={2}]{\otimes^{-1}} & \ctrl{1}
      & \gate{\rotatebox{270}{\textsc{ccm}$^{-1}$}} & \qw & \ \ldots\ \qw & \qw
      & \gate{\rotatebox{270}{\textsc{ccm}$^{-1}$}} & \qw
      & \rstick{$\ket{0}^{\otimes r}$}\qw \\
    \lstick{$\ket{m_i}$} & \qwbundle{k} &&&&&&&&&&& \gate{R} & \qw & \qw & \ \ldots\ \qw & \qw
      & \qw & \qw & \rstick{$\ket{n_i}$}\qw
  \end{tikzcd}
  \end{adjustbox}
  \caption{Circuit for \cref{alg:shear}. This diagram illustrates the computation $\ket{m_i} \gets \ket{n_i}$, using classically-controlled multiplication (CCM) for all arithmetic. The \framebox{$\otimes$} gate simply indicates the concatenation of two separate registers and requires no active operations. The \framebox{$R$} gate adds $+1$ to $n_i$ if $\ket{f}_{r-1} = \ket{1}$ (\ie, $f > \frac{1}{2}$), performing a rounding operation on $\ket{n_i}$.  The sign extension gate is defined in \cref{fig:qc_sign_notation}. Its notation reflects the fact that it is its own inverse. Since all multiplication is performed modulo the size of the product register, the second half of the circuit uncomputes the value stored in the $\ket{f}$ register only.}
  \label{fig:qc_shearing}
\end{figure*}

Two ancilla registers are required to perform the required arithmetic to sufficient accuracy.
The first is a register $\ket{f}$ appended to the coordinate being shifted in order to hold fractional coordinate values at intermediate steps.
The second is a register $\ket{e}$ used to extend the multiplicand $\ket{m_j}$ to properly account for the sign of the coordinate during multiplication~\cite{mano2004logic}.
At the end of the calculation only the integer part of the coordinate modulo $2^k$ is retained, with the ancilla registers reset by uncomputing their values.
In the algorithm and discussion below, we denote the register of the given coordinate extended by the fractional and sign extension ancilla registers above by $\ket{\tilde{n}_i}$ and $\ket{\tilde{m}_j}$, respectively.
In order to ensure that finite precision errors do not change the coordinate transformation by more than a single lattice step, both registers need to be at least of size
\begin{align}
    \label{eq:r_value}
    r = (k-1) + \ceil{\log_2(N-1)} \,.
\end{align}
The derivation of this result is presented in \cref{app:shearacc}.
Applying the shearing transformation around a non-integer point on the lattice only requires modifying the rounding step to apply a different offset.

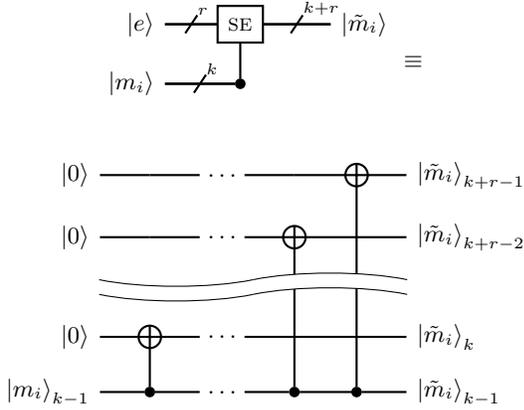
\begin{figure}[htb]
  \centering
    \begin{tikzcd}
      \lstick{$\ket{e}$}   & [2mm] \gate{\textsc{se}}\qwbundle{r} & [4mm] \rstick{\ket{\tilde{m}_i}}\qw\qwbundle{k+r} \\
      \lstick{$\ket{m_i}$} & [2mm] \ctrl{-1}\qwbundle{k} & \\
    \end{tikzcd} $\equiv$
    \medskip
    
    \begin{tikzcd}%[column sep= 0.2cm]
      \lstick{$\ket{0}$}         & \qw     & \ \ldots\ \qw
        & \qw     & \targ{} & \rstick{$\ket{\tilde{m}_i}_{k+r-1}$}\qw \\
      \lstick{$\ket{0}$}         & \qw     & \ \ldots\ \qw
        & \targ{} & \qw     & \rstick{$\ket{\tilde{m}_i}_{k+r-2}$}\qw \\
      \wave &&&&& \\
      \lstick{$\ket{0}$}         & \targ{} & \ \ldots\ \qw
        & \qw     & \qw     & \rstick{$\ket{\tilde{m}_i}_k$}\qw \\
      \lstick{$\ket{m_i}_{k-1}$} & \ctrl{-1}     & \ \ldots\ \qw
        & \ctrl{-3}     & \ctrl{-4}     & \rstick{$\ket{\tilde{m}_i}_{k-1}$}\qw
    \end{tikzcd}

  \caption{Sign extension sub-circuit, necessary for signed multiplication to be performed correctly. $\ket{m_i}_{k-1}$ is the sign qubit of $\ket{m_i}$. While not manifest in the compact notation adopted for clarity, sign extension is idempotent.}
  \label{fig:qc_sign_notation}
\end{figure}

A quantum circuit implementing this shearing procedure for a single coordinate is presented in \cref{fig:qc_shearing,fig:qc_sign_notation}.
The unitary operation implementing sign extension (\cref{fig:qc_sign_notation}) is its own inverse, and its uncomputation is distinguished only for conceptual clarity.
Since the multiplicand is unchanged by the multiplication, it returns $\ket{e}$ to its initial state.
The uncomputation of the CCM shifts \emph{only} the fractional register $\ket{f}$ and has the dual effect of performing necessary garbage collection and returning the fractional register to its initial state, allowing it to be reused in the shift of the next coordinate.
Beyond the $Nk$ state qubits, we require $2r+1$ ancillae, consisting of the $\ket{e}$ and $\ket{f}$ extension registers and 1 additional ancilla used as a carry during addition to minimize CNOT gates~\cite{Cuccaro_2004}.

\subsection{Resource requirements}

The gate cost of the resulting circuit is quadratic in $N$.
In particular, the total number of CNOT gates required is bounded by
\begin{align}
  \label{eq:shearing_cnots}
  \norm{\text{CNOT}} &\leq (N^2-N) \nnl
                     &\qquad (4k^2 + 8r^2 + 8kr + 26r + 11k - 8) \, ,
\end{align}
which is derived in \cref{app:shearcost}, 
The exact count is determined by the number of 1s in the binary representation of $M$ due to the use of CCM for arithmetic.
The typical number of CNOTs would be expected to be approximately half of this bound.

Even without implementing the KW 1D state preparation method, applying the shearing transformation to a set of 1D states prepared using the exponentially scaling algorithm of \cite{Klco:2019xro} yields a scaling of $\ord(N^2 k^2, 2^{k-1})$, compared to $\ord(2^{Nk})$ for a standard arbitrary state preparation.
If the number of qubits per lattice site $k$ is small, then $2^k$ can be treated as a fixed cost compared to $N^2$.

\begin{figure}[tbp]
    \includegraphics[width=0.9\columnwidth]{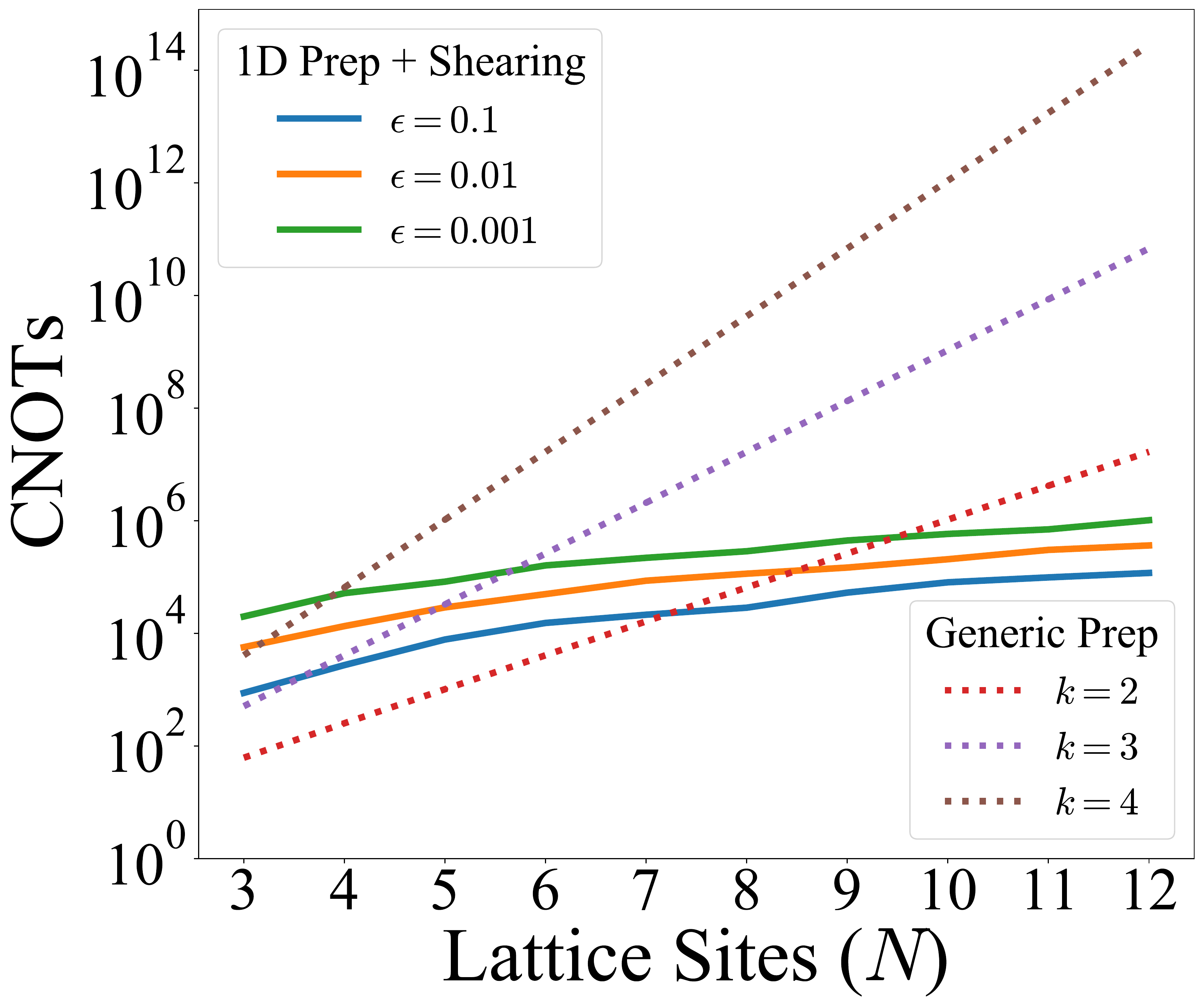}
    \caption{CNOT counts comparison: 1D state preparation plus shearing vs. standard state preparation. For the former, states that satisfy $1 - F(\psi_\text{shear},\psi_\text{opt})\leq \epsilon$ can be produced, given $k \sim \log N$. The latter produces optimal digitized states exactly, so the gate count is shown for relevant values of $k$.}
    \label{fig:cnots_shearing_vs_default}
\end{figure}

While the efficiency of the shearing transform is clear, we must also consider its precision. 
Analogously to the 1D case, we compute the state fidelity between states prepared using shearing and the optimal correlated state given by \cref{eq:NDoptstate}.
To understand how this fidelity behaves for states too large to simulate with classical resources, we also analytically approximate the resulting sum as an integral and by treating the average error due to rounding as an unknown parameter.
The resulting fidelity parametrically takes the form
\begin{align}
  \label{eq:fidapprox}
  F(\psi_\text{shear},\psi_\text{opt})
    \sim e^{\frac{2Nb}{\sigma^2}} \left(1 - \erf \frac{a}{\sigma}\right)^{2N} \, ,
\end{align}
where $\sigma$ is some characteristic width of the state in units of the lattice spacing and $a$ and $b$ are unknown constants $\lesssim 1$.
As the value of $b$ is only sensitive to subleading corrections at large width, in that limit the fidelity can be simplified to
\begin{align}
  F(\psi_\text{shear},\psi_\text{opt}) \sim e^{-\tilde{a} N/\sigma}
    \qquad \frac{N}{\sigma} \text{ fixed}, \sigma \to \infty \, ,
\end{align}
with $\tilde{a} = 4a/\sqrt{\pi}$.
This gives a simple $N/\sigma$ scaling to the expected error when this parameter is small.

Evidently, increasing the dimension of the multivariate Gaussian incurs a penalty in this fidelity.
However, this effect can be counteracted by modifying the width of the state in units of the lattice spacing. 
Let us choose $\sigma \sim 2^{k/2}$, as is the case for the physical scaling of simple harmonic oscillator we used in our 1D examples.
If the goal is to ensure the fidelity does not fall below some threshold, a 1D state register size scaling as $k \sim \log(N)$ will maintain a roughly constant fidelity. 
The resulting growth in the states being sheared implies $\ord(N^2\log^2 N)$ CNOTs will be required as $N$ is increased, with a commensurate $\ord(\log N)$ growth in ancillae.

This scaling is illustrated in \cref{fig:cnots_shearing_vs_default}, plotting CNOT count against $N$.
There is a large freedom in choosing the parameters of a generic multivariate Gaussian state, and to have a meaningful comparison across many values of $N$, we prepared states corresponding to the ground states of a (1+1)D scalar field theory on $N$ lattice sites, whose correlation matrices are reviewed in \cite{Jordan:2011ne}.
We prepare a product state of independent 1D Gaussians (\cref{eq:indep_gaussians}), then apply the shearing operator to obtain a correlated state.
Even for minimal states with $k = 2$, the shearing transform is more efficient for $N \ge 7$ dimensions compared to a generic real wavefunction preparation algorithm.
As $k$ is increased, generic exponential state preparation mechanisms become prohibitive for smaller values of $N$, with the shearing transform always more efficient once $k \ge 5$.

It should be noted that with the scaling above, an exponential-scaling 1D state algorithm amounts to just linear scaling in $N$.
Therefore, while for large enough $k$, it might become preferable to switch to a 1D prepartion algorithm polynomial in $k$, for the ultimate preparation of a multivariate Gaussian state, such a change will always be subleading to the $\ord(N^2)$ entangling gates required by the shearing transform itself.

\begin{figure}[htbp!]
    \subfloat[Fixed $N$, variable $k$]{
	    %\label{subfig:k1}
	    \includegraphics[width=0.9\columnwidth]{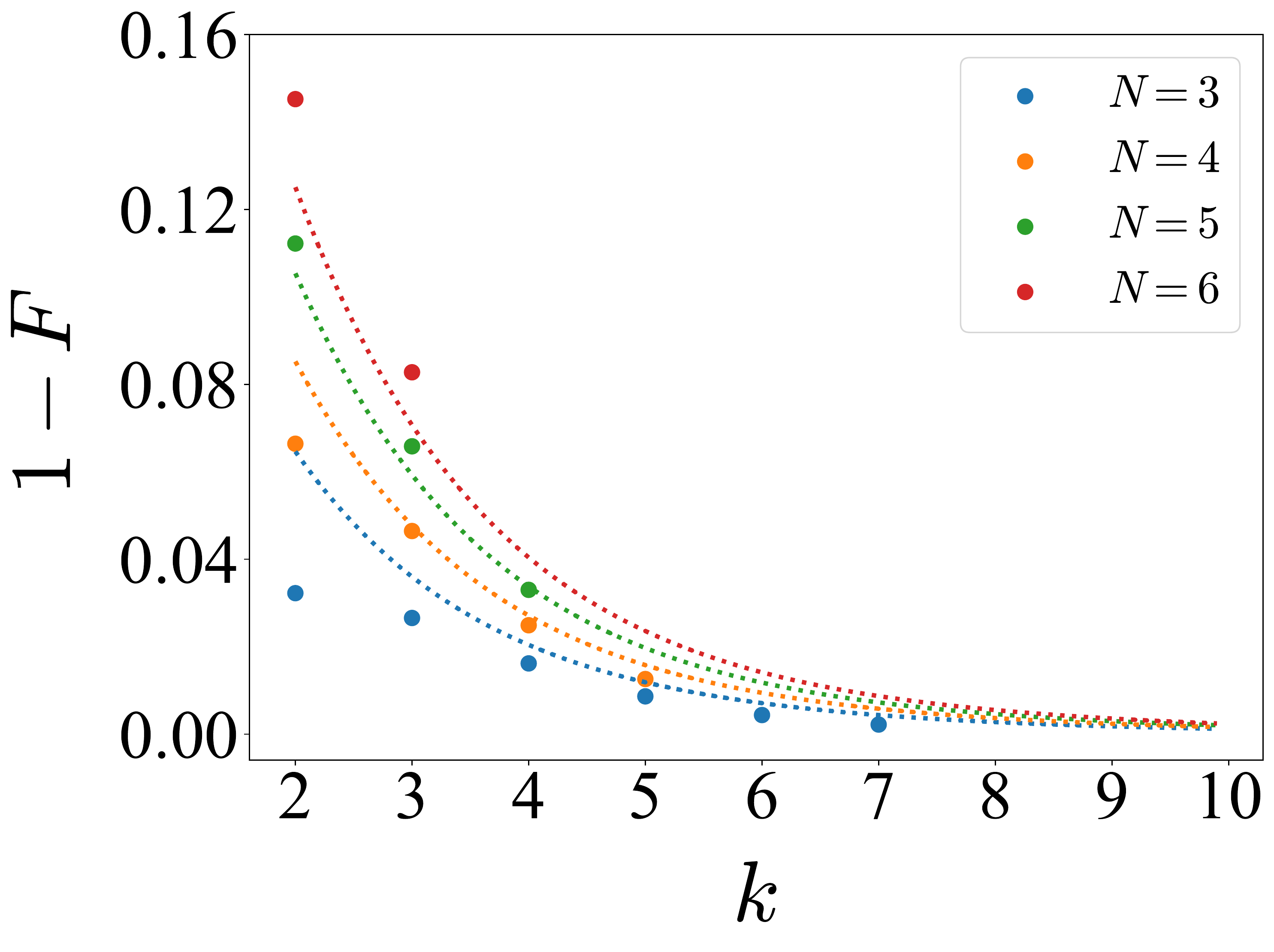}
	} 
	
	\subfloat[Fixed $k$, variable $N$]{
	    %\label{subfig:k3}
	    \includegraphics[width=0.9\columnwidth]{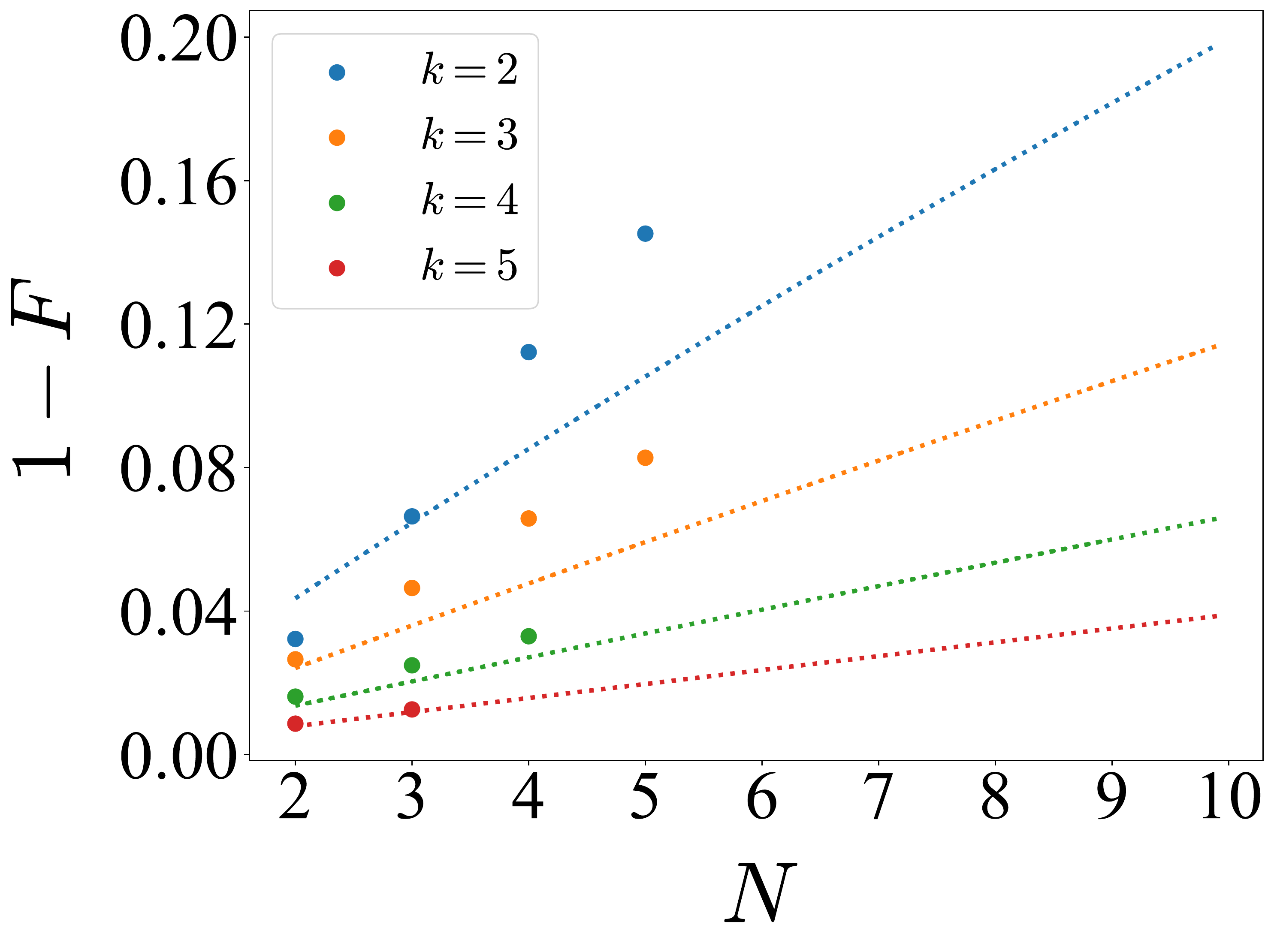}
	} 

	\caption{Fidelity comparison: Shearing vs. Optimal State Preparation. Data points represent fidelities computed using the \textsc{Qiskit} statevector simulator. The dashed lines are projections of the approximation \cref{eq:fidapprox}, with parameters fit to the computed data. The approximation becomes successively better as the resolution at which the state is sampled ($k$) increases.}
	\label{fig:fidelity_comp}
\end{figure}

\section{Summary and Outlook}
\label{sec:conc}

The efficient preparation of multivariate Gaussian states is likely to be a key ingredient in the quantum simulation of real-time dynamics for quantum field theory and many-body problems.
After a thorough review, we provided a set of algorithms for the implementation of Kitaev and Webb's multivariate Gaussian state preparation algorithm~\cite{kitaev2009wavefunction}, highlighting the accuracy and resource requirements of its two parts: preparing 1D Gaussian states and correlating dimensions via shearing transformations. 
Explicit quantum circuits implementing both parts of the algorithm, with detailed technical discussion going beyond the main text are provided in \cref{app:computealpha,app:CCMA}. 
These circuits offer systematically improvable approximations of the optimal digitized statevectors, as measured by state fidelity, while retaining polynomial resource scaling in state dimensionality $N$ and per-dimension register size $k$.
With explicit circuits, we can compute qubit and entangling gate counts for our circuit implementations to evaluate whether the asymptotic efficiency of the KW method can or should be realized on NISQ era devices.

The preparation of 1D Gaussians using polynomial-scaling methods requires a significant overhead due to the complexity of the quantum circuits required to calculate the superpositions of qubit rotations.
As illustrated in \cref{fig:cnots_1dprep}, simpler exponential-scaling algorithms, such as those of~\cite{Klco:2019xro}, require fewer resources unless states of $k \gtrsim 16$ qubits are required.
The KW algorithm also requires using a substantial number of additional ancillary qubits for arithmetic, as detailed in \cref{app:computealpha}.
However, future improvements to quantum arithmetic algorithms (addition and multiplication) may reduce the cost of the 1D KW algorithm, lowering the derived threshold at which the algorithm becomes less costly than an exponential one.
Even so, with energy eigenvalues known to converge to their continuum values exponentially in $k$, it may initially appear that exponential 1D state algorithms will satisfy most needs for the foreseeable future.

The application of a shearing transformation to $N$ 1D Gaussian states in order to create a correlated multivariate Gaussian has a lower CNOT cost for all but the smallest states.
For example, as shown in \cref{fig:cnots_shearing_vs_default}, already for $k = 4$, shearing is more efficient than a generic algorithm for $N \ge 4$, and for larger states shearing is always more efficient.
However, the approximations made to implement a shearing transform mean that fidelity of such states decreases linearly with $N$.
This loss of fidelity can compensated increasing the width of the state in units of the lattice spacing, with constant fidelity requiring $\sigma \sim N$.
If the earlier physical criterion for dimension digitization is adopted, such that $\sigma^2 \sim 2^k$, the overall cost of the shearing procedure to maintain fixed fidelity will require $\ord(N^2 \log^2 N)$ entangling gates, due to the fact that the size of the 1D input Gaussians needs to increase as $\ord(\log N)$ as well.
Likewise, the number of qubits required to implement shearing becomes $\ord(N\log N)$, as opposed to $Nk$ for the exponential state preparation.

The need to preserve multivariate state fidelity introduces a secondary need for increasing $k$ independent of the accuracy of the 1D states themselves.
For sufficiently large choices of $N$, a polynomially-scaling 1D state algorithms will be preferred for the preparation of the 1D inputs as well.
However, this regime is only approached logarithmically in $N$ and we estimate that while resources to create states with $N = \ord(100)$ dimensions are unavailable, which is likely to be the case for the majority of the NISQ era, direct preparation of 1D states will be preferable.
The $\ord(N^2 \log^2 N)$ scaling of the shearing operation means that in most cases of either regime, the cost of preparing 1D Gaussians is negligible compared to the cost of the shearing operation.

The need to preserve state fidelity with rising dimensionality will eventually call for all aspects of multivariate Gaussian state preparation to be handled by polynomial-scaling methods.
However, for many applications of interest and in the forseeable future, exponentially scaling algorithms to prepare a set of independent 1D Gaussians followed by the KW shearing transformations will yield the least resource intensive procedure for accurately constructing a multivariate correlated Gaussian state on a quantum computer.

\begin{acknowledgments}
We would like to thank Aniruddha Bapat for comments on the draft. CWB, PD, and BN are supported by the U.S. Department of Energy (DOE), Office of Science under contract DE-AC02-05CH11231. In particular, support comes from Quantum Information Science Enabled Discovery (QuantISED) for High Energy Physics (KA2401032) and the Office of Advanced Scientific Computing Research (ASCR) through the Accelerated Research for Quantum Computing Program. MF is supported by the DOE under grant DE-SC0010008. %This research used resources of the Oak Ridge Leadership Computing Facility, which is a DOE Office of Science User Facility supported under Contract DE-AC05-00OR22725.
\end{acknowledgments}

\clearpage

%%%%%%%%%%PLATO%%%%%%%%%%%%%%%

\appendix

\begin{widetext}

\section{Explicit examples of \texorpdfstring{\cref{eq:recursive}}{recursive state definition}}
\label{app:recursionex}

To build intuition for the recursive definition of the 1D Gaussian state presented in \cref{eq:recursive}, we work out the simplest cases explicitly, indicating the basis states in terms of their component qubits.
For $k=1$, the definitions of \cref{eq:cands} directly yield,
\begin{align}
    \ket{\xi_{\mu,\sigma;1}}
      = \sqrt{\frac{f\left(\frac{\mu}{2},\frac{\sigma}{2}\right)}{f(\mu,\sigma)}} \ket{0}
        + \sqrt{\frac{f\left(\frac{\mu-1}{2},\frac{\sigma}{2}\right)}
                      {f(\mu,\sigma)}} \ket{1}
      = c(\mu,\sigma) \ket{0} + s(\mu,\sigma) \ket{1} \, .
\end{align}
The first non-trivial result occurs for $k=2$, where 
\begin{align}
    \ket{\xi_{\mu,\sigma;2}}
      &= \sqrt{\frac{f\left(\frac{\mu}{4},\frac{\sigma}{4}\right)}{f(\mu,\sigma)}} \ket{00}
         + \sqrt{\frac{f\left(\frac{\mu-1}{4},\frac{\sigma}{4}\right)}
                      {f(\mu,\sigma)}} \ket{01}
         + \sqrt{\frac{f\left(\frac{\mu-2}{4},\frac{\sigma}{4}\right)}
                      {f(\mu,\sigma)}} \ket{10}
         + \sqrt{\frac{f\left(\frac{\mu-3}{4},\frac{\sigma}{4}\right)}
                      {f(\mu,\sigma)}} \ket{11} \nnl
      &= \sqrt{\frac{f\left(\frac{\mu}{4},\frac{\sigma}{4}\right)}
                    {f\left(\frac{\mu}{2},\frac{\sigma}{2}\right)}}
           \sqrt{\frac{f\left(\frac{\mu}{2},\frac{\sigma}{2}\right)}{f(\mu,\sigma)}} \ket{00}
         + \sqrt{\frac{f\left(\frac{\mu-1}{4},\frac{\sigma}{4}\right)}
                      {f\left(\frac{\mu-1}{2},\frac{\sigma}{2}\right)}}
             \sqrt{\frac{f\left(\frac{\mu-1}{2},\frac{\sigma}{2}\right)}{f(\mu,\sigma)}} \ket{01} \nnl
      &\qquad + \sqrt{\frac{f\left(\frac{\mu-2}{4},\frac{\sigma}{4}\right)}
                           {f\left(\frac{\mu}{2},\frac{\sigma}{2}\right)}}
                  \sqrt{\frac{f\left(\frac{\mu}{2},\frac{\sigma}{2}\right)}{f(\mu,\sigma)}} \ket{10}
              + \sqrt{\frac{f\left(\frac{\mu-3}{4},\frac{\sigma}{4}\right)}
                           {f\left(\frac{\mu-1}{2},\frac{\sigma}{2}\right)}}
                  \sqrt{\frac{f\left(\frac{\mu-1}{2},\frac{\sigma}{2}\right)}{f(\mu,\sigma)}}
                  \ket{11} \nnl
      &= c \left( \frac{\mu}{2}, \frac{\sigma}{2} \right) c(\mu,\sigma) \ket{00}
         + c \left( \frac{\mu-1}{2}, \frac{\sigma}{2} \right) s(\mu,\sigma) \ket{01}
         + s \left( \frac{\mu}{2}, \frac{\sigma}{2} \right) c(\mu,\sigma) \ket{10}
         + s \left( \frac{\mu-1}{2}, \frac{\sigma}{2} \right) s(\mu,\sigma) \ket{11} \nnl
      &= \left[ c \left( \frac{\mu}{2}, \frac{\sigma}{2} \right) \ket{0}
                + s \left( \frac{\mu}{2}, \frac{\sigma}{2} \right) \ket{1} \right]
           \otimes c(\mu,\sigma) \ket{0}
         + \left[ c \left( \frac{\mu-1}{2}, \frac{\sigma}{2} \right) \ket{0}
                  + s \left( \frac{\mu-1}{2}, \frac{\sigma}{2} \right) \ket{1} \right]
             \otimes s(\mu,\sigma) \ket{1} \, .
\end{align}
Eliding the intermediate steps in the derivation, we can similarly write
\begin{align}
  \ket{\xi_{\mu,\sigma;3}}
    &= \sum_{n=0}^7
          \sqrt{\frac{f\left(\frac{\mu-n}{8},\frac{\sigma}{8}\right)}{f(\mu,\sigma)}} \ket{n} \nnl
    &= \bigg\{ \left[ c \left( \frac{\mu}{4}, \frac{\sigma}{4} \right) \ket{0}
                      + s \left( \frac{\mu}{4}, \frac{\sigma}{4} \right) \ket{1} \right]
                 \otimes c \left( \frac{\mu}{2}, \frac{\sigma}{2} \right) \ket{0} \nnl
    &\qquad + \left[ c \left( \frac{\mu-2}{4}, \frac{\sigma}{4} \right) \ket{0}
                     + s \left( \frac{\mu-2}{4}, \frac{\sigma}{4} \right) \ket{1} \right]
                \otimes s \left( \frac{\mu}{2}, \frac{\sigma}{2} \right) \ket{1} \bigg\}
                \otimes c(\mu,\sigma) \ket{0} \nnl
    &\quad + \bigg\{ \left[ c \left( \frac{\mu-1}{4}, \frac{\sigma}{4} \right) \ket{0}
                             + s \left( \frac{\mu-1}{4}, \frac{\sigma}{4} \right) \ket{1} \right]
                        \otimes c \left( \frac{\mu-1}{2}, \frac{\sigma}{2} \right) \ket{0} \nnl
    &\qquad \quad + \left[ c \left( \frac{\mu-3}{4}, \frac{\sigma}{4} \right) \ket{0}
                     + s \left( \frac{\mu-3}{4}, \frac{\sigma}{4} \right) \ket{1} \right]
                \otimes s \left( \frac{\mu-1}{2}, \frac{\sigma}{2} \right) \ket{1} \bigg\}
                \otimes s(\mu,\sigma) \ket{1} \, .
\end{align}
\end{widetext}

\section{Quantum Arithmetic Circuits}
\label{app:qmathsummary}

There is a rich literature on quantum implementations of the core low-level operations, addition/subtraction and multiplication, which we summarize in \cref{tab:basiccomp}.
There is less existing work on more complicated operations such as division and square root, and the discussion of other transcendental functions is presently limited to implementations of series approximations. 
In order to provide the fairest accounting of ancilla and gate requirements in \cref{alg:1Dprep}, our goal is to construct the most efficient quantum algorithm for computing the rotation angle.

The fundamental building block of all arithmetic functions is the adder. 
Three main categories of quantum adders have been studied and optimized: ripple-carry~\cite{Cuccaro_2004, Thapliyal_2013, takahashi2009quantum, Li_2020}, carry-lookahead~\cite{draper2004_claa, thapliyal2020_claa}, and quantum Fourier transform (QFT)-based~\cite{draper2000_qft, florio2004_qft, Ruiz_Perez_2017}. 
The QFT-based adders are strictly worse than ripple-carry adders in terms of resource cost, and therefore our choice is between ripple-carry and carry-lookahead adders.
Carry-lookahead adders have increased entangling gate and ancilla costs, to the benefit of decreased delay (longest chain of entangling gates that must be executed in series), and since presently gate errors are typically more limiting than the total coherence time, we use the ripple adder of~\cite{Cuccaro_2004} in our resource counts going forward.

\begin{table}[tb]
  \begin{tabular}{llll}
    \toprule
                    & Gate Count  & Delay          & Ancilla   \\
    adders          &             &                &           \\
    \midrule
    ripple-carry~\cite{Cuccaro_2004, Thapliyal_2013, takahashi2009quantum, Li_2020}
                    & $\ord(n)$   & $\ord(n)$      & 0 or 1    \\
    carry-lookahead~\cite{draper2004_claa, thapliyal2020_claa}
                    & $\ord(n)$   & $\ord(\log n)$ & $\ord(n)$ \\
    QFT-based~\cite{draper2000_qft, florio2004_qft, Ruiz_Perez_2017}
                    & $\ord(n^2)$ & $\ord(n)$      & 0        \\ [1.0ex]
    % \midrule
    multipliers     &             &                &           \\
    \midrule
    shift-and-add~\cite{munozcoreas2017_mult}
                    & $\ord(n^2)$ & $\ord(n^2)$    & $\ord(n)$ \\
    QFT-based~\cite{Ruiz_Perez_2017, florio2004_qft, Zhang_2020}
                    & $\ord(n^3)$ & $\ord(n^2)$    & $\ord(n)$ \\
    Karatsuba~\cite{Kowada:jucs_12_5,parent2017_karatsuba}
                    & $\ord(n^{1.58})$ & $\ord(n^{1.16})$ & $\ord(n^{1.43})$  \\
    \bottomrule
  \end{tabular}
  \caption{Basic arithmetic quantum circuits. Gate count and delay (longest gate series in circuit) are given by expressing all entangling operations in terms of CNOT gates, while $n$ is the qubit-length of operands.}
  \label{tab:basiccomp}
\end{table}

Quantum multipliers broadly split into two categories: analogues of classical shift-and-add (S/A) algorithms~\cite{munozcoreas2017_mult}, and QFT-based algorithms~\cite{Ruiz_Perez_2017, florio2004_qft, Zhang_2020}. 
As with addition, QFT-based multipliers have worse gate count than S/A multipliers, and can be ignored for our purposes.
Variants of S/A multiplication taking advantage of recursive schemes can be more efficient asymptotically, both in classical and quantum implementations.
As a particular example, Karatsuba multiplication offers improved scaling with both gate count and depth, at the cost of worse scaling with ancillae to implement the recursion~\cite{Kowada:jucs_12_5,parent2017_karatsuba}.
However, the prefactors in the scaling of such recursive schemes are typically substantial and at present circuits only become more efficient for multiplication of numbers requiring $n > \ord(1000)$ qubits to encode.
Given that qubits will be at a premium on NISQ devices, we use a fixed-precision version of the S/A multiplier given in~\cite{munozcoreas2017_mult} for our comparison circuits in this paper.
This algorithm cannot store the result in place, as information from both inputs is needed to compute each partial sum, and the product is ultimately stored in an additional register.

\addtolength{\tabcolsep}{8pt}
\begin{table*}[tp]
  \begin{adjustbox}{width=\textwidth}
    \begin{tabular}{lllll}
        \toprule
          & Gate Count              & Gate Delay    & Ancilla & Notes       \\
        \midrule
        non-restoring division~\cite{thapliyal2018_division}       
          & $18n^2 + \ord(n)$       & $10n^2 + \ord(n)$           & $n-1$   & integer quotient + remainder\\
        division (Newton's method)~\cite{bhaskar2015_division}    
          & $33dn^2 + \ord(dn)$     & $27dn^2 + \ord(dn)$         & $(d+2)n$      & \\
        \midrule
        non-restoring square root~\cite{munozcoreas2018_sqrt}
          & $\frac{9}{2}n^2 + \ord(n)$ & $5n^2+\ord(n)$ & $n+1$   & integer root + remainder\\
        square root (Newton's method)~\cite{haner2018optimizing}
          & $55dn^2 + \ord(dn)$     & $45dn^2 + \ord(dn)$         & $(d+4)n$        & \\
        \midrule
        polynomial (series approximation)
          & $11dn^2 + \ord(dn)$     & $9dn^2 + \ord(dn)$          & $dn$           & \\
        parallel polynomial \cite{haner2018optimizing}
          & $11dn^2 + \ord(dn)$     & $9dn^2 + \ord(dn)$          & $dn + \log \ell$    & \\
        \bottomrule
    \end{tabular}
  \end{adjustbox}
  \caption{Circuit costs of more advanced arithmetic circuits. Gate count and delay (longest gate series in circuit) are given by expressing all entangling operations in terms of CNOT gates. Subcircuits include ripple adders \cite{Cuccaro_2004} and multipliers \cite{munozcoreas2017_mult}. Here $n$ is the qubit-length of operands while $d$ represent either the polynomial degree or number of iterations when using Newton's method. The parameter $\ell$ denotes the number of sets of coefficients to evaluate in parallel.}
  \label{tab:complexcomp}
\end{table*}
\addtolength{\tabcolsep}{-8pt}

Like their classical counterparts, other arithmetic operations are generally more costly to implement compared to addition and multiplication.
\Cref{tab:complexcomp} summarizes gate and qubit costs of current state-of-the-art circuits implementing division, square root, and polynomial evaluation necessary for series approximations.
Since the trigonometric and Jacobi theta functions (or exponential functions if using the summation series definition of \cref{eq:f}) would need to be evaluated via series approximation, computing all of the functions appearing in the definition of the rotation angle would induce substantial computational costs, and would make tracking the resulting accuracy harder, compared to simply working with a given series approximation of \cref{eq:alpha} directly.
However, we cannot discount the possibility that future improvements of quantum arithmetic algorithms may significantly reduce overhead costs for more complicated functions, favoring their use over the piecewise series approximations we employ below.

\section{\boldmath Computing \texorpdfstring{$\ket{\alpha(\mu_j,\sigma_j)}$}{the rotation angle}}
\label{app:computealpha}

Implemented directly as written, the evaluation of $\alpha(\mu_j,\sigma_j)$ defined by \cref{eq:alpha} requires a somewhat involved list of arithmetic functions.
Beyond addition and multiplication, the operations of division, square root, and inverse cosine are necessary.
Alternatively, the original KW suggestion of directly evaluating the sum in \cref{eq:f} involves computing the exponential of multiple terms as an approximation.
If a quantum circuit capable of computing $\alpha_j$ to the desired accuracy for arbitrary values of $\mu_j$ and $\sigma_j$ needed to be implemented, we would require an efficient quantum implementation of each step above.

Luckily, this is not necessary.
At each recursion level $\sigma_j$ takes a unique value and as discussed in \cref{sec:registers}, for a judicious choice of indexing, we can always ensure $-1 < \mu_j < 0$.
Therefore, we only need quantum circuits capable of evaluating $\alpha(\mu_j,\sigma_j)$ for fixed values of $\sigma_j$ and constrained to the range $\mu_j \in (-1,0)$.

\subsection{Approximating \texorpdfstring{$\alpha(\mu,\sigma)$}{the rotation angle}}
\label{app:approxalpha}

The full functional dependence of $\alpha(\mu,\sigma)$ is rather complex, but since $\sigma_j = \sigma/2^j$ is a fixed parameter at each recursion level we are free to select the most appropriate approximation of the dependence on $\mu_j$ for a given choice of $\sigma_j$.
As a function of $\mu_j$, $\alpha(\mu,\sigma)$ is always a function of period 2 with no simple rapidly-converging universal approximation.
However, in the limits of very large and very small $\sigma_j$, it can be well approximated by simpler functions.
For large $\sigma_j$, \cref{eq:alpha,eq:f_as_theta} lead to $\tilde{\alpha}(\mu_j,\sigma_j)$ having a sinusoidal dependence on $\mu_j$ around an average value of $\frac{1}{2}$ whose amplitude is exponentially suppressed by $\sigma_j$.
For small $\sigma_j$, $\tilde{\alpha}(\mu_j,\sigma_j)$ is exponentially close to a square wave of unit amplitude.
For intermediate regions with $\sigma_j \sim \ord(1)$, the general functional behavior is complex, with sinusoidal variation growing with decreasing $\sigma_j$ until it saturates to the above square wave behavior.

For our purposes, the periodicity of $\alpha(\mu,\sigma)$ is not relevant.
Since we need to evaluate the angle for $\mu_j \in (-1,0)$, we only need the fact that on this interval it is an odd function centered on $\mu_j = -\frac{1}{2}$.
With this in mind, we parameterize $\tilde{\alpha}(\mu_j,\sigma_j)$ by the following form,
\begin{align}
\label{eq:alphareg}
  \tilde{\alpha}(\mu_j,\sigma_j) =
    \begin{cases}
      \tilde{\alpha}^{(d)}_\text{hi}(\mu'_j,\sigma_j)
        &\text{for } \sigma_j \ge \sigma_\text{hi} \\
      \tilde{\alpha}^{(d,d')}_\text{int}(\mu_j,\sigma_j)
        &\text{for } \sigma_j \in (\sigma_\text{lo}, \sigma_\text{hi}) \\
      \Theta(-\mu'_j)
        &\text{for } \sigma_j \le \sigma_\text{lo} \\
    \end{cases} \, ,
\end{align}
where $\mu'_j = \mu_j + \frac{1}{2}$ and $\Theta(x)$ denotes the Heaviside step function.
In the large $\sigma_j$ region we use the above symmetry properties of $\alpha(\mu,\sigma)$ to approximate it as a $(2d+1)$th-order polynomial,
\begin{align}
\label{eq:alphahi}
  \tilde{\alpha}^{(d)}_\text{hi}(\mu'_j,\sigma_j)
    &= \frac{1}{2} + \sum_{n=0}^{d(b,\sigma_j)} a_n(\sigma_j) (\mu'_j)^{2n+1} \, .
\end{align}
The coefficients $a_n$ depend on $\sigma_j$, while the order of approximation $d$ also depends on the accuracy with which $\alpha_j$ can be stored in its $b$-qubit register.
For sufficiently large $\sigma_j$ and small $b$, $\tilde{\alpha}_\text{hi}$ takes the constant value $\frac{1}{2}$ making the calculation trivial.

\begin{figure}[b]
  \tikzmath{\ysize=3.5; \xsize=7;
            \mumid=1.1; \alim=0.5;}
  \centering
  \begin{tikzpicture}
    \draw[->,ultra thick] (0,0)--(-\xsize-0.4,0) node[left]{\large $\mu_j$};
      \draw (-\xsize,0.1)--(-\xsize,-0.1) node[below]{-1};
      \draw (-\xsize/2,0.1)--(-\xsize/2,-0.1) node[below]{$-\frac{1}{2}$};
      \draw (0,0.1)--(0,-0.1) node[below]{0};
    \draw[->,ultra thick] (0,0)--(0,4.4) node[above]{\large $\tilde{\alpha}_j$};
      \draw (-0.1,0)--(0.1,0) node[right]{0};
      \draw (-0.1,\ysize/2)--(0.1,\ysize/2) node[right]{$\frac{1}{2}$};
      \draw (-0.1,\ysize)--(0.1,\ysize) node[right]{1};
    \draw[dash dot, thick, color=gray!80] (-\xsize/2+\mumid,0)--(-\xsize/2+\mumid,\ysize);
    \draw[dash dot, thick, color=gray!80] (-\xsize/2-\mumid,0)--(-\xsize/2-\mumid,\ysize);
      \draw[<->] (-\xsize/2-\mumid, \ysize+0.2) -- (-\xsize/2+\mumid, \ysize+0.2)
        node[midway, above]{$2\mu_\text{mid}$};
    \draw[dotted, thick, color=gray!80] (-\xsize,\ysize-\alim)--(-\xsize/2-\mumid,\ysize-\alim)
      node[right, color=black]{$1-a'_0(\sigma_j)$};
    \draw[dotted, thick, color=gray!80] (0,\alim)--(-\xsize/2+\mumid,\alim)
      node[left, color=black]{$a'_0(\sigma_j)$};
    \draw[very thick, color=black!60!green, domain=-\xsize/2+\mumid:0]
      plot(-\xsize-\x, \ysize - \alim - 0.05*\x*\x);
    \draw[very thick, color=black!60!green, domain=-\xsize/2+\mumid:0]
      plot(\x, \alim + 0.05*\x*\x);
    \draw[very thick, color=black!40!red, domain=-\mumid:\mumid]
      plot(\x-\xsize/2, \ysize/2 - 1.15*\x + 0.25*\x*\x*\x);
    \draw[<-, thick, color=black!70!red!60] (-3.4,2.0) to [out=60, in=210](-1.25,2.6)
      node[above, align=center]{\small $(2d+1)^\text{th}$-order\\
                                              odd polynomial\\around $-\frac{1}{2}$};
    \draw[<-, thick, color=black!70!green!60] (-5.8,2.7) to [out=-85, in=90](-6.2,1.7)
      node[below, align=center]{\small $(2d')^\text{th}$-order\\even polynomial\\around -1};
  \end{tikzpicture}
  \caption{Visual summary of the interpolating function $\tilde{\alpha}^{(d,d')}_\text{int}$ defined in \cref{eq:alphaint}. The same form with $\mu_\text{mid} = \frac{1}{2}$ describes $\tilde{\alpha}^{(d)}_\text{hi}$.}
  \label{fig:alphaint}
\end{figure}
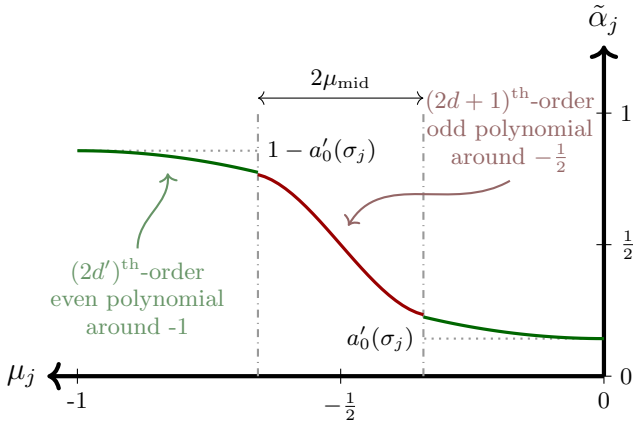

In the intermediate region, a rapidly converging series valid over the entire required range of $\mu_j$ does not exist, so we chose a piecewise polynomial that interpolates between the two regimes above.
Specifically,
\begin{align}
\label{eq:alphaint}
  \tilde{\alpha}^{(d,d')}_\text{int}
    &= \begin{cases}
         \tilde{\alpha}^{(d')}_\text{out}(\mu_j,\sigma_j)
           &\text{for } \mu'_j \ge \mu_\text{mid} \\
         \tilde{\alpha}^{(d)}_\text{hi}(\mu'_j,\sigma_j)
           &\text{for } |\mu'_j| < \mu_\text{mid} \\
         1 - \tilde{\alpha}^{(d')}_\text{out}(\mu''_j,\sigma_j)
           &\text{for } \mu'_j \le -\mu_\text{mid}
       \end{cases} \, ,
\end{align}
with $\mu''_j = \mu_j + 1$ and
\begin{align}
\label{eq:alphaout}
  \tilde{\alpha}^{(d')}_\text{out}(\mu_j,\sigma_j)
    = \sum_{n=0}^{d'(b,\sigma_j)} a'_n(\sigma_j)\, \mu_j^{2n} \, .
\end{align}
As $\sigma_j$ decreases, asymptotically $\tilde{\alpha}_\text{out}^{(d')}, \mu_\text{mid} \to 0$, while the slope of $\tilde{\alpha}_\text{hi}^{(d)}$ at the midpoint increases super-exponentially, providing the required interpolating behavior.
This parameterization is sketched in \cref{fig:alphaint}, while the various parameters appearing in the series approximation are extracted from the series expansion of the function around the mid- and end-points.

For our numerical studies, we chose the transition points $(\sigma_\text{lo}, \sigma_\text{hi}) = (0.0375, 0.6)$, such that the most complicated intermediate form is only evaluated 4 times independent of the number of qubits $k$ used to encode the state.
In general, $\sigma_\text{lo}$ also has a weak dependence on $b$ since the Heaviside step function must be sufficiently accurate to avoid the need for corrections in our parameterization, but as noted above this behavior is approached super-exponentially.
We find that keeping only 1 or 2 terms in the $a^{(\prime)}_n(\sigma_j)$ series is sufficient to avoid introducing additional errors beyond those due to angle discretization itself for $b \lesssim 10$ qubits.

For this approximation, we present a quantum circuit implementing the operation
\begin{align}
    \ket{0}^{\otimes b} \ket{\mu_j} \ket{0}^{\otimes \norm{\text{anc}}} 
      \to \ket{\tilde{\alpha}(\sigma_j,\mu_j)} \ket{\tilde{\mu}_j} \ket{a}.
\end{align}
This differs slightly from the form given in \cref{alg:1Dprep} due to the inclusion of ancillae and potential modification of the $\mu_j$ register --- and hence the state qubits.
While the latter change may appear to be cause for concern, all resulting changes are uncomputed at the the end of each recursive step.

\subsection{\texorpdfstring{$\sigma_j > \sigma_\text{hi}$: }{}Polynomial approximation}
\label{app:polycirc}

For $\sigma_j > \sigma_\text{hi}$, a single polynomial series, \cref{eq:alphahi}, provides a good approximation for $\tilde{\alpha}(\mu_j, \sigma_j)$.
The algorithm we present here for its evaluation also serves as an input to more complex case of $\sigma_j \in (\sigma_\text{lo}, \sigma_\text{hi})$ below.

Since in this case the polynomial is odd in $\mu'_j = \mu_j + \frac{1}{2}$, it is most efficient to perform this variable transformation directly in the state register.
This requires augmenting the $\ket{q_{j-1} \dotsi q_0}$ register with 2 qubits to store the least significant bit of \cref{eq:mu5_explicit} and the sign of $\mu_j$.
The transformation to $\mu'_j$ can then be performed in place.
For technical reasons we store $\ket{-\mu'_j}$ in the resulting register,\footnote{In two's complement binary, which we use for all subsequent signed arithmetic, preparing $\ket{-\mu_j}$ itself requires no entangling operations, while preparing $\ket{\mu_j}$ would use $\ord(j)$ CNOTs to change its sign due to the manner in which $\mu_j$ is stored in the $\ket{q_{j-1} \dotsi q_0}$ register. The CNOT cost of the subsequent shift to produce $\ket{\pm\mu'_j}$ is the same in both cases.} and apply compensating modifications to the evaluation of the polynomial where necessary.

The $d$th-order polynomial itself is efficiently evaluated using Horner's method~\cite{Burden81:numerical}.
After computing $(\mu'_j)^2$ and storing it in a dedicated register, we compute the following series of iterates,
\begin{align}
  x_1 &\equiv a_{d-1}(\sigma_j) + a_d(\sigma_j) \cdot (\mu'_j)^2 \, , \\
  x_2 &\equiv a_{d-2}(\sigma_j) + x_1 \cdot (\mu'_j)^2 \, , \\
      &\vdots \nnl
  x_d &\equiv a_0(\sigma_j) + x_{d-1} \cdot (\mu'_j)^2 \, .
\end{align}
The computational cost of this scheme is $d$ additions and $d$ multiplications.
For an even polynomial, this yields the desired answer, while for the odd polynomial series here, we conclude by computing $\frac{1}{2} - x_d \cdot (-\mu'_j)$ in the angle register.
This last step is the only one requiring compensation compensation for the sign of the $\ket{-\mu'_j}$ register, accomplished by subtracting rather than adding the partial sums during S/A multiplication.

\begin{figure*}[htp]
  \begin{adjustbox}{width=\textwidth}
    \begin{tikzcd}[column sep= 0.4cm]
      \lstick{$a_d$} & \cw & \cw & \cw & \cw \cwbend{3} \\
      \lstick{$\ket{0}^{\otimes b}$} & \qwbundle{b}
        & \qw & \qw & \qw & \ket{a_{d-1}} \qw \vqw{3} & \qw & \ket{a_{d-2}} \qw \vqw{4}
        & \qw & \ket{a_{d-3}} \qw \vqw{5} & \qw & \ \ldots\ \qw & \qw
        & \ket{a_0}\qw \vqw{6} & \gate{\times} \vqw{1} &\gate{+\frac{1}{2}}
        & \rstick{$\ket{\tilde{\alpha}(\mu_j,\sigma_j)}$} \qw \\
      \lstick{$\underbrace{\ket{-\mu_j}}_{\ket{0.q_{j-1}\dotsc q_01}}$} & \qwbundle{j+2}
        & \gate{-\frac{1}{2}} & \ctrl{1}
          \gategroup[wires=7,steps=11,style={dashed, rounded corners, fill=gray!15, inner xsep=2pt},
                     background, label style={label position=below,anchor=north,yshift=-0.2cm}]
                    {\large $U_\text{iter}$}
        & \qw & \qw & \qw & \qw & \qw & \qw & \qw & \ \ldots\ \qw  & \qw & \qw & \ctrl{5} & \qw
        & \rstick{$\ket{-\mu_j'}$} \qw \\
      \lstick{$\ket{0}^{\otimes b}$} & \qwbundle{b}
        & \qw & \gate{\square^2} & \ctrl{1} & \qw & \ctrl{1} & \qw & \ctrl{2}
        & \qw & \ctrl{3} & \ \ldots\ \qw  & \ctrl{4} & \qw & \qw & \qw
        & \rstick{$\ket{(\mu')^2}$} \qw \\
      \lstick{$\ket{0}^{\otimes(b+1)}$} & \qwbundle{b+1}
        & \qw & \qw & \gate{\rotatebox{270}{CCM}} & \gate{+} & \ctrl{1} & \qw
        & \qw & \qw & \qw & \ \ldots\ \qw  & \qw & \qw & \qw & \qw
        & \rstick{$\ket{x_1}$} \qw \\
      \lstick{$\ket{0}^{\otimes(b+1)}$} & \qwbundle{b+1}
        & \qw & \qw & \qw & \qw  & \gate{\times} & \gate{+} & \ctrl{1} & \qw
        & \qw & \ \ldots\ \qw  & \qw & \qw & \qw & \qw
        & \rstick{$\ket{x_2}$} \qw \\
      \lstick{$\ket{0}^{\otimes(b+1)}$} & \qwbundle{b+1}
        & \qw & \qw & \qw & \qw & \qw & \qw & \gate{\times} & \gate{+}
        & \ctrl{1} & \ \ldots\ \qw  & \qw & \qw & \qw & \qw
        & \rstick{$\ket{x_3}$} \qw \\
      \wave &&&&&&&&&&&&&&&& \\
      \lstick{$\ket{0}^{\otimes(b+1)}$} & \qwbundle{b+1}
        & \qw & \qw & \qw & \qw & \qw & \qw & \qw & \qw & \qw
        & \ \ldots\ \qw  & \gate{\times} \vqw{-1} & \gate{+} \vqw{-1} & \ctrl{-1} & \qw
        & \rstick{$\ket{x_d}$} \qw
    \end{tikzcd}
  \end{adjustbox}
  \caption{Circuit for computing $\tilde{\alpha}(\mu_j,\sigma_j)$ when $\sigma_j \ge \sigma_\text{hi}$. The {\small \framebox{$\times$}} gates denote the result of multiplying the two values stored in registers denoted by controls. The \framebox{$+$} gates add the connected coefficient $\ket{a}$ to the register to which \boxed{+} is applied, while the coefficient itself is encoded and erased from the control register before and after the operation. The \framebox{CCM} gate is the classically-controlled multiplication-accumulation circuit illustrated in \cref{fig:qc_ccma2}. The \framebox{$\square^2$} gate denotes the result of squaring the value stored in the register denoted by a control, which is nothing but a multiplication gate with both control registers identical. For all these gates, the control nodes are shorthand for inputs that do not get modified by the gate, and are not simple binary controls. Finally, the \framebox{$\pm\frac{1}{2}$} gates represent a constant shift by $\pm\frac{1}{2}$. The dashed box highlights the $U_\text{iter}$ multi-controlled unitary for computing the iterates in Horner's method. This unitary also appears in the calculation for $\sigma_j \in (\sigma_\text{lo}, \sigma_\text{hi})$ shown in \cref{fig:alpha_int_circ}, with the replacement of the CCM gate by a normal multiplication.}
  \label{fig:alpha_hi_circ}
\end{figure*}
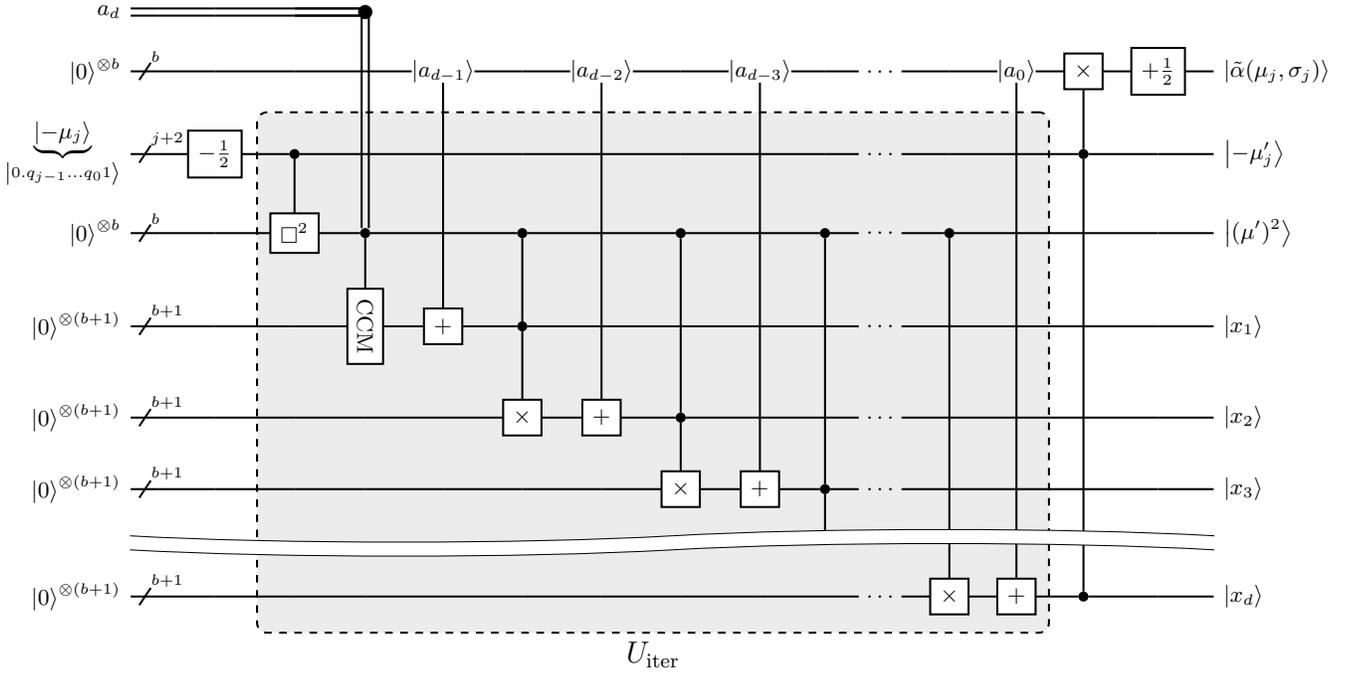

The number of qubits required to store intermediate computation results, like the order of the polynomial itself, is determined by the accuracy with which $\tilde{\alpha}_j$ can be stored in its $b$-qubit register.
To avoid introducing additional errors beyond those due to the truncation of $\tilde{\alpha}_j$, we require the total error of our approximation to satisfy $\log_2 \epsilon \le -(b+1)$.
This selects a value of $d$ given $\sigma_j$, since we need only keep terms for which $\log_2 a_n(\sigma_j) > 2n - b$.
The size of the final register is known, so all iterate ancilla registers can be chosen to have size $b+1$, without introducing additional truncation errors.
The additional qubit compared to $\tilde{\alpha}_j$ occurs because although the rotation angle is always positive, the iterates computed at intermediate stages may not be.
Further register size reductions are possible if the numerical values of coefficients are known to be sufficiently small since $|\mu'_j|^2 < \frac{1}{4}$, but are not employed here.
The details of the quantum circuit that implements this computation are presented as an algorithm below and as a circuit diagram in \cref{fig:alpha_hi_circ}. 

\begin{algorithm}[H]
  \DontPrintSemicolon
  \KwData{$\ket{\mu_j} \sim \ket{q_{j-1} \dotsi q_0}$, angle precision $b$, polynomial order $d(b,\sigma_j)$, series coefficients $a_{0, \dotsc, d}(\sigma_j)$}
  \KwResult{state $\ket{\tilde{\alpha}(\mu,\sigma)} \otimes \ket{a} \otimes \ket{-\mu'_j}$}
  \Begin{
    Initialize registers:\; \Indp
    mean register $\ket{-\mu_j} \gets \ket{0} \otimes \ket{q_{j-1} \dotsi q_0} \otimes \ket{1}$\;
    $(\mu'_j)^2$ register $\ket{(\mu'_j)^2} \gets \ket{0}^{\otimes b}$\;
    iterate registers $\ket{x_1}, \dotsc, \ket{x_d} \gets \ket{0}^{\otimes (b+1)}$\;
    angle register $\ket{\tilde{\alpha}} \equiv \ket{x_{d+1}} \gets \ket{0}^{\otimes (b+1)}$\; \Indm
    Compute arguments of polynomial:\; \Indp 
    $\ket{-\mu'_j} \gets \ket{-\mu_j} - \frac{1}{2}$\;
    $\ket{(\mu'_j)^2 = 0} \gets \ket{-\mu'_j} \cdot \ket{-\mu'_j}$\; \Indm
    $\ket{x_1 = 0} \gets a_d \cdot \ket{(\mu'_j)^2}$ \tcp*[r]{via CCM (\cref{app:CCMA})}
    $\ket{\tilde{\alpha} = 0} \gets \ket{a_{d-1}}$\;
    $\ket{x_1} \gets \ket{\tilde{\alpha} = a_{d-1}}
                     + \ket{x_1 = a_d \cdot (\mu'_j)^2}$\;
    $\ket{\tilde{\alpha} = a_{d-1}} \gets \ket{0}^{\otimes b}$\;
    \For{$n \gets 2$ \KwTo $d$}{
      $\ket{x_n = 0} \gets \ket{x_{n-1}} \cdot \ket{(\mu'_j)^2}$\;
      $\ket{\tilde{\alpha} = 0} \gets \ket{a_{d-n}}$\;
      $\ket{x_n} \gets \ket{\tilde{\alpha} = a_{d-n}} + \ket{x_n = x_{n-1} \cdot (\mu'_j)^2}$\;
      $\ket{\tilde{\alpha} = a_{d-n}} \gets \ket{0}^{\otimes b}$\;
    }
    $\ket{\tilde{\alpha}} \gets -(\ket{x_d} \cdot \ket{-\mu'_j})$\;
    $\ket{\tilde{\alpha} = x_d \cdot \mu'_j} \gets \ket{\tilde{\alpha} = x_d \cdot \mu'_j} + \frac{1}{2}$;
  }
  \caption{Odd polynomial for $\tilde{\alpha}(\mu_j,\sigma_j)$}
  \label{alg:alpha_poly}
\end{algorithm}

In addition to the state register implicitly storing $\mu_j$ and the angle register itself, the number of ancilla needed to perform the computation is
\begin{align}
   \norm{\text{anc}} &= (d+1)b + d + 2 \,,
\end{align}
where $d$ is itself dependent on $b$, leading to an overall ancilla scaling of $\ord(b \log b)$.
The gate counts in CNOTs is:
\begin{align}
 \begin{split}
    \norm{\text{CNOT}} \leq \,\,& 11(4bj - 2j^2 + db^2) - 3b^2 \\
    &\quad - 13j + (32d + 39)b \\
    &\quad + 10d - 56\,.
 \end{split}
 \label{eq:cnots_highsigma}
\end{align}
The number of ancilla qubits required could be reduced by using ``pebbling'' strategies~\cite{haner2018optimizing}, at the cost of a significantly higher gate count.

\subsection{\texorpdfstring{$\sigma_j \in (\sigma_\text{lo}, \sigma_\text{hi})$: }{}Piecewise functions}
\label{app:piececirc}

For smaller values $\sigma_j < \sigma_\text{hi}$, the convergence of any one series expansion for all required values of $\mu_j$ begins to involve an increasingly large number of terms with large cancellations term-by-term.
The resulting calculation would both require increasingly large resources and be subject to increasingly large numerical instabilities.
Instead, for $\sigma_j \in (\sigma_\text{lo}, \sigma_\text{hi})$ the piecewise approximation of \cref{eq:alphaint} becomes preferable.

To efficiently evaluate this expression, we use a modified version of the parallel polynomial evaluation technique presented in~\cite{haner2018optimizing} to evaluate the polynomials
\begin{align}
\label{eq:multipoly}
    \sum_{n=0}^{\max(d,d')} a^{(\prime)}_n \left( \mu_j^{(\prime,\prime\prime)} \right)^{2n} \, ,
\end{align}
in parallel.
Here the polynomials are a function of $\mu_j$, $\mu'_j$, or $\mu''_j$ depending on which region in \cref{eq:alphaint} is relevant, with the coefficients likewise set for the appropriate to the region.
In the case where $d' \ne d$, the unneeded coefficients for the lower-order polynomial are simply set to 0.
This is accomplished by using a separate label register to encode the choice of region for all values of $\mu_j$ in parallel.
All subsequent aspects of the calculation are then conditioned on this register value.

For the label, we chose the encoding
\begin{align}
\label{eq:partition}
     \ket{\ell} = \begin{cases} 
                        \ket{00} &\text{for } \mu' > \mu_\text{mid}\\
                        \ket{10} &\text{for } |\mu'| < \mu_\text{mid}\\
                        \ket{01} &\text{for } \mu' < -\mu_\text{mid}
                  \end{cases} \, .
 \end{align}
This register plays a triple role in the subsequent circuit.
First, a controlled shift is performed in the initial mean/state register $\ket{-\mu_j}$ to set the correct polynomial argument.
Second, polynomial coefficients in the arithmetic circuit are also encoded in a controlled manner.
(Note that with our choice of labelling scheme, this second operation is controlled solely by the first qubit of the label register.)

The polynomial circuit presented in \cref{app:polycirc} can be straightforwardly modified to produce the a linear combinations of entangled label and iterate registers.
In particular with respect to the final iterate, the controlled polynomial circuit will produce a linear combination of
\begin{align}
  \ket{\ell} \otimes\ket{x_d} =
     \begin{cases}
       \ket{0y} \otimes \ket{\sum_{n=0}^{d'} a'_n(\sigma_j) (\mu^{(\prime\prime)}_j)^{2n}} \\
       \ket{10} \otimes \ket{\sum_{n=0}^d a_n(\sigma_j) (\mu'_j)^{2n}}
     \end{cases}
\end{align}
states for $y$ taking either value.

There is one additional step needed to arrive at $\tilde{\alpha}_\text{int}(\mu_j,\sigma_j)$ of \cref{eq:alphaint}. 
The iterate entangled with $\ket{\ell = 10}$ must be transformed into $\frac{1}{2} + x_d \cdot \mu'_j$ as before, while that entangled with $\ket{\ell = 01}$ must be transformed into $1 - x_d$.
Both of these, along with the identity transformation for $\ket{\ell = 00}$ are implemented in parallel by the calculation of $s_1 + s_2 x_d$ in an additional register with coefficients again controlled by the label register.
The only complication compared to the circuits we've already presented is that for $\ket{\ell = 10}$, $b_1$ is not a known constant but instead must be initialized via a controlled copy from the mean/state register.

\begin{figure*}[tb]
  \centering
    \begin{tikzcd}[column sep= 0.4cm]
      \lstick{$\ket{0}^{\otimes 2}$}      & \qwbundle[alternate=2]{} 
        & \gate[nwires=1]{\text{Label}}\qwbundle[alternate=2]{} &  \ctrlbundle[2]{2} & \ctrlbundle[2]{1} & \ctrlbundle[2]{1} &  \ \ldots\ \qwbundle[alternate=2]{} & \ctrlbundle[2]{1}
        & \ctrlbundle[2]{2} & \qwbundle[alternate=2]{} & \ctrlbundle[2]{1}
        & \rstick{$\ket{\ell}$} \qwbundle[alternate=2]{} \\
      \lstick{$\ket{0}^{\otimes b}$}      & \qwbundle{b} 
        & \qw & \qw & \ket{a^{(\prime)}_d} \qw \vqw{1} & \ket{a^{(\prime)}_{d-1}} \qw \vqw{1}
        & \ \ldots\ \qw & \ket{a^{(\prime)}_0} \qw \vqw{1} & \qw & \gate{\times} & \gate{\Longstack{$\emptyset$ $+\frac{1}{2}$ +1}}
        & \rstick{$\ket{\tilde{\alpha}(\mu_j,\sigma_j)}$} \qw \\
      \lstick{$\ket{-\mu_j}$} & \qwbundle{j+2}
        & \ctrl{-2} & \gate{\Longstack{$\emptyset$ $+\frac{1}{2}$ +1}} & \qw\gategroup[wires=4, steps=4, style={inner xsep=0pt, inner ysep=0pt, fill=white}, label style={yshift=-1.8cm}]{\Large $U_\text{iter}$} & & & \qw & \ctrl{4} & \qw & \qw
        & \rstick{$\ket{-\mu^{(\prime, \prime\prime)}_j}$} \qw \\
      \lstick{$\ket{0}^{\otimes b}$}      & \qwbundle{b+1}
        & \qw & \qw & \qw & & & \qw & \qw & \qw & \qw
        & \rstick{$\ket{(\mu^{(\prime, \prime\prime)}_j)^2}$} \qw \\
      \lstick{$\ket{0}^{\otimes (d-1)(b+1)}$} & \qwbundle[alternate]{}
        & \qwbundle[alternate]{} & \qwbundle[alternate]{}
        & \qwbundle[alternate]{} & & & \qwbundle[alternate]{}
        & \qwbundle[alternate]{} & \qwbundle[alternate]{} & \qwbundle[alternate]{}
        & \rstick{$\ket{x_1} \otimes \dotsi \otimes \ket{x_{d-1}}$} \qwbundle[alternate]{} \\
      \lstick{$\ket{0}^{\otimes (b+1)}$}      & \qwbundle{b+1}
        & \qw & \qw & \qw & & & \qw & \qw & \ctrl{-4} & \qw
        & \rstick{$\ket{x_d}$} \qw \\
      \lstick{$\ket{0}^{\otimes (b+1)}$}      & \qwbundle{b+1}
        & \qw & \qw & \qw & \qw & \ \ldots\ \qw & \qw & \gate{\Longstack{$\ket{+1}$ $\ket{\mu'_j}$ $\ket{-1}$}} & \ctrl{-1} & \qw
        & \rstick{\ket{s_2}} \qw
    \end{tikzcd}
    \caption{Circuit for computing $\tilde{\alpha}(\mu,\sigma)$ when $\sigma_j\in(\sigma_{\text{lo}},\sigma_{\text{hi}})$. The additional and multiplication gates function as in \cref{fig:alpha_hi_circ}, where the $U_\text{iter}$ multi-controlled unitary is also defined. The notation for the control wires going into $U_\text{iter}$ is shorthand for a label-controlled encoding of the coefficient in the angle register, the coefficient acting as an arithmetic parameter as in \cref{fig:alpha_hi_circ} for $U_\text{iter}$, and a controlled erasure of the coefficient from the angle register. When appearing side by side, the encoding and erasure of the coefficients can be combined to reduce the gate count, as is done in our estimates in the text. The \framebox{Label} gate is a generic circuit for preparing label register $\ket{\ell}$ that requires $\ord(j)$ gates. The final two label-controlled gates implement the controlled linear transformation of $x_d$ to transform it into the angle in each piecewise region.}
    \label{fig:alpha_int_circ}
\end{figure*}
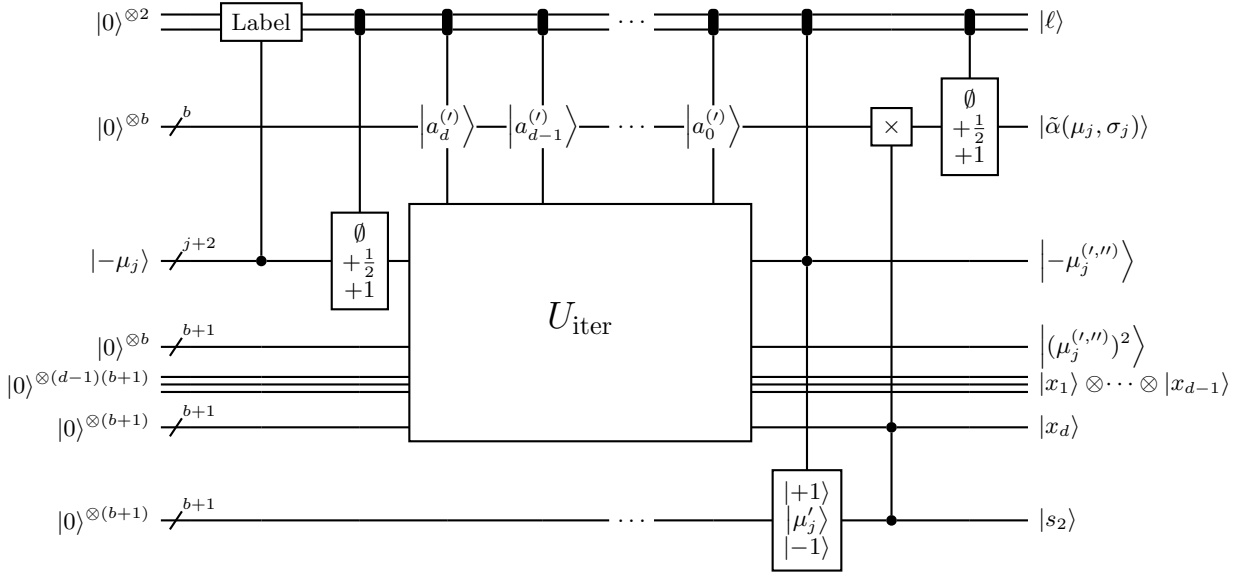

The resulting circuit is presented in \cref{fig:alpha_int_circ}.
In addition to state register implicitly storing $\mu_j$ and the angle register itself, the number of ancilla needed to implement this circuit is:
\begin{align}
    \norm{\text{anc}} &= 3 + b + d(b+1) + (j+2)\,,
\end{align}
and the number of CNOTs used is:
\begin{align}
 \begin{split}
    \norm{\text{CNOT}} \leq& 11[4bj - 2j^2 + db^2] \\
    &\quad + [34d + 90]b - 7j + \ord(j) \\
    &\quad + 10d - 76\,,
 \end{split}
 \label{eq:cnots_intsigma}
\end{align}
where as before, $d(b, \sigma_j)$ is dependant on both $\sigma_j$ and the desired accuracy of the final answer.
If a degree $d=1$ series is used, the gate count is reduced by $11b^2 + 34b + 10$ CNOTs.
As in the $\sigma_j>\sigma_{\text{hi}}$ region, the number of ancilla qubits required can be reduced using ``pebbling'' strategies \cite{haner2018optimizing}, in exchange for a significantly higher gate count.

\subsection{Heaviside step function implementation}
\label{app:stepfun}

The evaluation at very small values of $\sigma$ is much simpler.
Explicitly rewriting the relevant case,
\begin{align}
\label{eq:alpha_slo}
  \tilde{\alpha}(\mu_j,\sigma_j \le \sigma_\text{lo}) =
    \begin{cases}
      0 &\text{if } \mu_j > -0.5 \\
      1 &\text{if } \mu_j < -0.5
    \end{cases} \, .
\end{align}
If $\mu = -\frac{1}{2}$, $\mu_j \ne -\frac{1}{2}$ for all $j > 0$ so we never need to specify the value of the angle at the transition value.

\begin{figure}[tb]
    \centering
    \begin{tikzcd}
        \lstick{$\ket{q_{j-1}}$} & \ctrl{1}        & \rstick{$\ket{q_{j-1}}$} \qw \\
        \lstick{$\ket{q_j}$}     & \gate{R_y(\pi)} & \rstick{$\ket{q_j}$}     \qw
    \end{tikzcd}
    \caption{Implicit computation of $\alpha_j$ and rotation $\ket{q_j}$ for $\sigma \le \sigma_\text{lo}$.}
    \label{fig:alpha_lo_circ}
\end{figure}
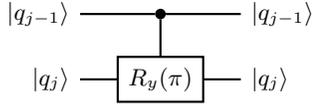

In this case, computing $\alpha_j$ and rotating $\ket{q_j}$ can be combined as follows:
Given the encoding of $\mu_j$ described in \cref{sec:registers}, the condition $\mu_j > -0.5$ ($\mu_j < -0.5$) can be restated as $q_{j-1} = 0$ ($q_{j-1} = 1$).
Therefore, no rotation is applied to $\ket{q_j}$ if $\ket{q_{j-1}} = \ket{0}$, while $R(\pi/2) \equiv R_y(\pi)$ is applied to $\ket{q_j}$ if $\ket{q_{j-1}} = \ket{1}$.
The explicit computation of $\alpha_j$ can then be bypassed and a controlled rotation applied directly to $\ket{q_j}$, as illustrated by \cref{fig:alpha_lo_circ}.
No uncomputation is necessary in this case, since no angle or state registers are changed, and the entire recursive step costs only 2 CNOTs.
In fact, since before this step $\ket{q_j}$ is always in the $\ket{0}$ state, we can replace $R_y(\pi)$ with $X$ to achieve the same effect, reducing the cost to a single CNOT.

\section{Classically controlled multiplication}
\label{app:CCMA}

\begin{figure}[htp]
  \begin{adjustbox}{width=\columnwidth}
    \begin{tikzcd}
      \lstick{$M$}       & \cwbend{1}\cw        & \cwbend{1}\cw & \ \ldots\ \cw & \cwbend{1} & \\
      \lstick{$\ket{m}$} & \ctrl{1}\qwbundle{k} & \ctrl{2}      & \ \ldots\ \qw & \ctrl{4}
                         & \rstick{$\ket{m}$}\qw \\
      \lstick{$\ket{n}_0$} & \gate[4,disable auto height,nwires={3}]{\rotatebox{270}{$RA(k)$}}
                           & \qw & \ \ldots\ \qw & \qw
                           & \rstick{$\ket{n \oplus Mm}_0$}\qw \\
      \lstick{$\ket{n}_1$} && \gate[3,disable auto height,nwires={2}]{\rotatebox{270}{$RA(k-1)$}}
                           & \ \ldots\ \qw & \qw
                           & \rstick{$\ket{n \oplus Mm}_1$}\qw \\
      \vdots \hspace{1cm}  &&& \vdots && \hspace{1cm} \vdots \\
      \lstick{$\ket{n}_k$} &&& \ \ldots\ \qw & \gate{RA(1)} & \rstick{$\ket{n \oplus Mm}_k$}\qw \\  \\
    \end{tikzcd}
  \end{adjustbox}
  \caption{Classically-controlled multiplication (CCM) circuit, with $\oplus$ indicating addition modulo the size of register $\ket{n}$. Here, classical control means that the $j^\text{th}$ adder (in our implementation the ripple adder of~\cite{Cuccaro_2004}) is inserted into the circuit if the $j^\text{th}$ (from least to most significant) bit in the binary representation of $M$ is 1, otherwise it is omitted. \framebox{$RA(\ell)$} denotes the $\ell$-qubit adder adding the control to the target. As we are interested in the modular addtion of $n \oplus Mn$, when $\ell < k$, we simply ignore the highest $k-\ell$ qubits of $\ket{m}$. The modification to fixed-precision non-integer arithmetic, as required by our algorithms, merely requires the appropriate offsets to the adders, with some adders potentially only using a subset of the most-significant qubits of $\ket{m}$.}
  \label{fig:qc_ccma2}
\end{figure}
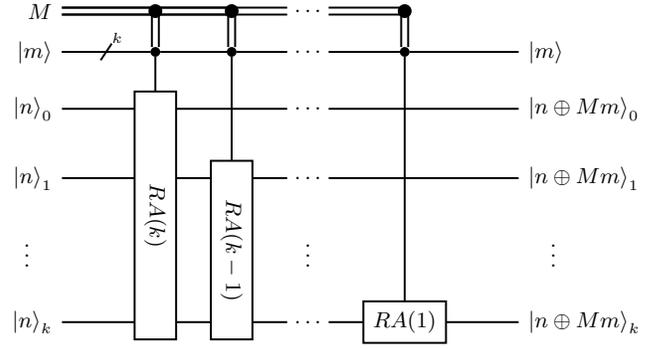

Here, we present a particular implementation of the classically controlled multiplication (CCM) algorithm described in \cref{sec:shearcircuit}.
In this case, we accumulate products of the form $M_{ij}m_j$ where $M_{ij}$ is a classical binary number with $k$ integral bits and $r$ fractional bits, and $m_j$ is a ($k$-bit integer) $(k+1)$-bit half integer stored on a fixed-point $(k+r)$-qubit register.
Additionally, we only need the result modulo $2^k$, so we use modular arithmetic.
Therefore, \cref{alg:ccma} details CCM under these restrictions, although it is easily generalized to non-modular, floating-point arithmetic with non-integer $m_j$.

In our implementation of the shearing operation given by \cref{eq:quantum_shearing_trafo}, the computational basis $\ket{m_j}$ states must be mean-centered.
Interpreting these basis states as integers in two's complement, this means that if our 1D Gaussian states are prepared symmetrically, the ``mean-centered basis states'' are half-integers rather than integers.
For example, the computational basis states of a $k$-qubit system in two's complement notation are $[-2^{k-1}, ... , 0, ... , 2^{k-1}-1]$.
On this discrete set, a symmetric Gaussian has mean $-\frac{1}{2}$, so the mean-centered basis states are the half-integers $[-2^{k-1}+ 1/2, \dotsc , 1/2, \dotsc , 2^{k-1} - 1/2]$.
Therefore, we implement CCM using a classical multiplier $M$ and half-integer quantum multiplicand $\ket{n+1/2}$.
To represent $\ket{n+1/2}$, just append one fractional qubit in the $\ket{1}$ state to $k$-qubit integer $\ket{n}$.
A qubit from the $\ket{e}$ register (see \cref{fig:qc_shearing}) can be reused for this purpose, so CCM with half-integers has the same qubit cost as CCM with integer multiplicands.

\begin{algorithm}[tbp]
\vspace{4pt}
 \textbf{Definitions:}\\[8pt]
 Let quantum register $\ket{n}$ store a $k$-bit integer.\\[8pt]
 Let quantum register $\ket{n_0}$ store a $(k+r)$-bit number, and define the concatenation $\ket{n_0} = \ket{n_0}_{\text{int}}\ket{n_0}_{\text{dec}}$. Register $\ket{n_0}_{\text{int}}$ stores the $k$ integral bits of $\ket{n_0}$ and register $\ket{n_0}_{\text{dec}}$ stores the $r$ fractional bits. \\[8pt]
 Let $M$ be classical number, represented in binary using $k$ integral bits and $r$ fractional bits.\\[8pt]
Note: All binary representation use two's complement notation.\\[8pt]
 \KwResult{$\ket{n_0}\ket{n} \, \mapsto \, \ket{n_0+M(n+1/2)}\ket{n}$}\vspace{8pt}
 \textbf{Algorithm:}\\[8pt]
 Initialize an additional $r$-qubit register:
 \begin{enumerate}
     \item[] $\ket{e}=\ket{0},\,$ ``extension'' register
 \end{enumerate}
 Copy the sign qubit of $\ket{n}$ to each qubit of $\ket{e}$, using $r$ CNOTs total. Therefore, the concatenation $\ket{e}\ket{n}$ and $\ket{n}$ represent the same two's complement number, but $\ket{e}\ket{n}$ is padded by $r$ extra integral qubits. \\[8pt]
 (First partial sum:) If the least significant bit ($M[0]$) of $M$ is `1', apply a quantum addition circuit that adds $\ket{e}\ket{n}$ to $\ket{n_0}_{\text{int}}\ket{n_0}_{\text{dec}}$. \\[8pt]
 Uncopy the sign qubit of $\ket{n}$ from $\ket{e}_{r-1}$, using a CNOT. Then set $\ket{e}_{r-1}=\ket{1}$. This functions as the $1/2$-place qubit of $\ket{n+1/2}$. \\[8pt]
 \For{$j= 1$ \textbf{to} $r$}{
    \begin{enumerate}
        \item[] If the $j$-th least significant bit ($M[j]$) of $M$ is `1', apply a quantum addition circuit that adds $\ket{e}_{0..r-j}\ket{n}\ket{e}_{r-1}$ to $\ket{n_0}_{\text{int}}\ket{n_0}_{\text{dec}}$. This is equivalent to adding $n+1/2$, right-shifted by $r-j$ bits, to $n_0$.\\[8pt]
        Note that this addition circuit has size $k+1+r-j$.
    \end{enumerate}
 }
 \For{$j= r+1$ \textbf{to} $k+r-1$}{
    \begin{enumerate}
        \item[] If the $(j)$-th least significant bit ($M[j]$) of $M$ is `1', apply a quantum addition circuit that adds $\ket{e}_{r-1}\ket{n}_{0..k+r-j}$ to $\ket{n_0}_{\text{int},\:j-r-1..k-1}$. This is equivalent to adding $n+1/2$, left-shifted by $j$ bits, to $n_0$. \\[8pt]
        Note that this addition circuit has size $k+1+r-j$.
    \end{enumerate}
 }
 \vspace{10pt}
 Un-copy the sign qubit of $\ket{n}$ from $\ket{e}_{0..r-2}$, using $r-1$ CNOTs total.
 \caption{CCM algorithm for modular, fixed-point arithmetic.}
 \label{alg:ccma}
\end{algorithm}

The extension register $\ket{e}$ is necessary to implement the quantum addition circuits in \cref{alg:ccma}.
Starting with $j=0$, $k$-bit integer $\ket{m}$ is added to $k$ bits of $\ket{\tilde{n}}$, with the least significant bit of $\ket{n}$ shifted to line up with the $j^\text{th}$ least significant bit $\ket{\tilde{n}}$.
This computation is performed in place, storing the result on quantum register $\ket{\tilde{n}}$.
Therefore, to carry the addition through $\ket{\tilde{n}}$, $k+r-j$ carry bits must be computed.
However, to compute all $k+r-j$ carry bits, the register storing $k$-bit integer $\ket{n}$ must also contain $(k+r-j)$ bits.
This is implemented by copying the sign qubit of $\ket{m}$ to $r$ extension qubits (using CNOTs) appended to the left of $\ket{m}$.
(As noted in \cref{alg:ccma}, this does not change the value of $\ket{m}$ in two's complement notation.)
Then, $r-j$ of these extension bits are used to implement the $j$-th addition circuit.
\Cref{fig:qc_ccma2} illustrates the circuit diagram of the CCM algorithm, and \cref{tab:ccma_cnot} its CNOT gate cost.

\begin{table*}[tbp]
 \begin{tabular}{ll}
  \toprule
    Step & $\norm{\text{CNOT}}$ \\
  \midrule
    extend sign of $\ket{m}$ to $\ket{e}$                 & $r$                       \\[6pt]
    if $M[0] = 1$, apply a ripple adder$(k+r)$            & $16(k+r)$                 \\[6pt]
    uncopy $\ket{e}_{r-1}$                                & $1$                       \\[6pt] 
    $\forall j = 1, k+r-1$, if $M[j] = 1$, apply a ripple ddder$(k+1+r-j)$
      & $\sum_{j=1}^{k+r-1} \left[ 16(k+1+r-j) \right]$ \\[6pt]
    uncopy sign of $\ket{m}$                              & $r-1$                     \\
  \midrule
    total
      & $\leq 8(k^2 + r^2 + 2kr + 3k + 3r - 2) + 2r$ \\
  \bottomrule
 \end{tabular}
 \caption{CNOT count for \cref{alg:ccma}. $M[j]$ is the $j^\text{th}$ least significant bit of classical number $M$. The true number of CNOTs used depends on the value of $M$, while the table entry is an upper bound computed by assuming all $M[j]=1$.}
 \label{tab:ccma_cnot}
\end{table*}

\section{Determining the precision of the shearing matrix}
\label{app:shearacc}

Although each $\ket{m_i}$ is an integer, the off-diagonal shearing matrix entries $M_{ij}$ are decimal numbers that must be approximated by binary representations with $r$ fractional bits.
The question is how many fractional bits are necessary to compute $\ket{n_i}$ such that the approximation error is minimized.
Let $N$ be the dimensions of the Gaussian state, $k$ the number of qubits encoding each dimension, and $r$ the number of fractional qubits used to specify $M_{ij}$ and compute
\begin{align}
    n_i = \quad m_i + \sum_{j=i+1}^{N-1} M_{ij} m_j \,.
\end{align}
This approximation induces an absolute error in $M_{ij}$ of at most $2^{-(r+1)}$.
Therefore the maximum possible absolute error incurred computing $n_i$ is
\begin{align}
    \epsilon_i(r) &= \frac{1}{2^{r+1}} \sum_{j=i+1}^{N-1}|n_j| \,.
\end{align}
Then, given that each $n_j \in \{-2^{k-1},...,2^{k-1}-1\}$,
\begin{align}
    \epsilon_i(r) &\leq \frac{1}{2^{r+1}} \sum_{j=i+1}^{N-1} 2^{k-1} \nnl
                     &= (N-i-1)\, 2^{k-r-2} \,.
\end{align}

Because at the end of the coordinate transformation, only the part of $\ket{n_i}$ rounded to the nearest integer is retained, minimizing the approximation error really means making sure that the approximate and exact values of $n_i$ round to the same number.
Unfortunately, it isn't possible to ensure zero error.
For example, if the exact and approximate values of $n_i$ lie at $(1 \pm \epsilon)/2$, the two will round to different integers for arbitrarily small error.
However, we can ensure that the approximate value is within 1 lattice spacing of the exact value, which is true if $\epsilon_i(r) < \frac{1}{2}$.
This restriction implies
\begin{align}
    r > (k-1)+\log_2(N-i-1) \, .
\end{align}
The strongest restriction comes from computing the coordinate transformation of $\ket{n_0}$.
It thus suffices to choose
\begin{align}
    r = (k-1) + \ceil{\log_2(N-1)} \,
\end{align}
to ensure rounding errors never shift any coordinate by more than one lattice spacing.
Larger values of the register will reduce the possibility of such a shift without ever bringing it to zero.

\section{Shearing CNOT costs}
\label{app:shearcost}

\addtolength{\tabcolsep}{8pt}
\begin{table*}[tbp]
 \begin{tabular}{ll}
    \toprule
    Step                                                    & $\norm{\text{CNOT}}$ \\
    \midrule
    applying CCM to add $M_{ij}m_j$ to $n_i$
      & $\sum_{j=i+1}^{N-1} \left[ \leq 8(k^2 + r^2 + 2kr + 3k + 3r - 2) + 2r \right]$ \\[6pt]
    applying CCM$^{-1}$ to uncompute $\ket{f}$
      & $\sum_{j=i+1}^{N-1} \left[ \leq 8(r^2+3r) + 2(r-k) \right]$ \\
    \bottomrule
 \end{tabular}
 \caption{Gate count of \cref{alg:shear}. Each step is applied for all values of $j$ from $i+1$ to $N-1$. $M_{ij}[n]$ represents the $n^\text{th}$ least significant bit of the binary representation of $M_{ij}$, with $r$ fractional qubits. Note that the extension qubits are also used for uncomputation, so $2r$ gates are omitted from the second summand.}
 \label{tab:shearing}
\end{table*}
\addtolength{\tabcolsep}{-8pt}

Here we derive the entangling gate cost of implementing the shearing transformation on a quantum computer. 
First, gate counts of the different steps of \cref{alg:shear}, which is used to compute $\ket{n_i}$, are summarized in \cref{tab:shearing}.
Adding up these subcounts, an upper bound on the total number of CNOTs required to compute $\ket{n_i}$ is

\begin{align}
  &\sum_{j=i+1}^{N-1} \left[ 2(2r - k) + 8(k^2 + 2r^2 + 2kr + 3k + 6r - 2) \right] \nnl
  &\quad = (N-i-1) \nnl
  &\quad \qquad (8k^2 + 16r^2 + 16kr + 52r + 22k - 16)
  \label{eq:shearing_gates_2}
\end{align}

Finally, we compute $\ket{n_0},\dots,\ket{n_{N-2}}$ in sequence, so the total CNOT count for the shearing operation is given by summing over $i$ from 0 to $N-2$,
\begin{align}
  \norm{\text{CNOT}}
    &= \sum_{i=0}^{N-2} \text{\cref{eq:shearing_gates_2}} \nnl
    &= (N^2-N) \nnl
    &\qquad (4k^2 + 8r^2 + 8kr + 26r + 11k - 8)
\end{align}

\bibliographystyle{apsrev4-2}
\bibliography{KWimplement}

\end{document}